\documentclass[journal=tches,final]{iacrtrans}

\usepackage{lipsum}
\usepackage{tikz}
\usepackage{amsmath,amssymb,amsfonts}
\usepackage{graphicx}
\usepackage{xurl}
\usepackage{subfigure}
\usepackage{bm}

\usepackage{makecell}
\usepackage{algorithm, algorithmic}
\usepackage{array} 
\usepackage{caption}
\usepackage{graphbox}
\usepackage{todonotes}
\usepackage{threeparttable}
\usepackage{colortbl}
\usepackage{framed}
\usepackage{multirow}
\usepackage{pifont}
\usepackage {indentfirst} 
\usepackage{booktabs}
\usepackage{tabularx}
\usepackage{hyperref}
\usepackage{bm}
\usepackage{ragged2e}
\newcolumntype{C}{>{\centering\arraybackslash}X}

\newcommand{\btw}{\texttt{Bit2Watt}}

\author{Zhouhao Ji, Kaikai Pan\thanks{Corresponding author.} \and Wenyuan Xu}
\institute{
  Zhejiang University, Hangzhou, China \\
  \email{zhouhao0779@zju.edu.cn, pankaikai@zju.edu.cn, wyxu@zju.edu.cn}
}

\title[\texttt{Bit2Watt}]{\texttt{Bit2Watt}: A Cyber–Physical Vulnerability Exploiting GPU Workloads Across Power and Computing Infrastructures}

\begin{document}

\maketitle


\begin{abstract}
Modern data centers increasingly rely on large-scale GPU clusters and on-site renewable energy resources, resulting in a tightly coupled cyber–physical system between computing workloads and power-electronic-dominated grids. In this paper, we reveal \btw, a previously unexplored vulnerability in which an adversary manipulates GPU workloads to induce controlled, high-frequency power modulations that destabilize local power infrastructure and propagate back to disrupt computing services.
Unlike traditional attacks that compromise grid-side devices or communication channels, \btw~operates entirely within the cyber layer as a legal tenant, which could amplify fluctuations, harmonic distortion, and damping degradation, particularly in high-DER-penetration scenarios. 
This risk is difficult to detect under routine cloud- and facility-side monitoring because \btw~exploits legitimate workload execution paths and concentrates much of its distinctive behavior in high-frequency components that are weakly captured by common telemetry.
We validate \texttt{Bit2Watt} through impedance-based analysis, power system simulations, and real-world experiments on GPUs and grid-connected PV inverters. 
Under the synchronized worst-case aggregation model studied in the paper, manipulating 1,000 GPUs in a 1-MW local power system with 90\% DERs raises current THD to 46.8\% and results in a damping ratio of -0.27. We further show that the resulting power-quality degradation can stress data-center power-delivery equipment, trigger protection mechanisms, and, in extreme simulated cases, induce cascading failures in transmission-scale systems. In addition, we analyze a plausible \texttt{Watt2Bit} feedback path, including denial-of-service risks and covert information exfiltration via EMI side channels.
This work exposes a vulnerability at the intersection of modern computing and renewable-integrated power systems, highlighting the urgent need for cross-layer defenses that jointly consider workload scheduling and power electronics.

\end{abstract}

\section{Introduction}
Modern digital infrastructure is increasingly built upon the deep integration of large-scale computing systems and electrical power networks.
The rapid proliferation of GPU-accelerated workloads—driven by artificial intelligence and high-performance computing (HPC)—has dramatically increased both the magnitude and volatility of the power demand in data centers, which varies at sub-millisecond timescales. In parallel, power networks worldwide are transforming toward high penetration of inverter-control-based distributed energy resources (DERs), such as photovoltaics (PVs). 
As a result, modern data centers and renewable-integrated grids together constitute a networked cyber–physical system, in which cyber-level workload scheduling directly drives physical-layer actuation through power-electronic interfaces and fast control loops.

While prior research \cite{madiot2.0,blackiot,GridShock, GPUsidechannel,timingattack,SIDECHANNEL, cacheside} has independently highlighted the security issues in power systems and computing systems, existing cyber–physical threat models largely overlook adversaries who do not compromise sensing, control or communication infrastructure, but instead exploit legitimate cyber-level control inputs to induce cyber–physical security issues. Modern cloud platforms powered by data centers allow untrusted but fully legitimate tenants to control thousands of GPUs through official software, raising a critical question: \emph{\textbf{can purely computational actions, executed as legitimate workloads, be weaponized to destabilize power infrastructure?}} This question motivates our study, as such risks would bypass conventional perimeter defenses by operating entirely within legitimate access privileges and exploiting the intrinsic cyber–physical coupling of next-generation infrastructures.

Addressing this question is promising yet challenging. First, revealing the principle of how to induce power fluctuation through GPU tasks is a challenge. Second, it remains challenging to rigorously analyze how computation-induced power modulations propagate and interact with the dynamics of power infrastructure. To overcome the challenges, we first propose two kinds of task-based power modulation methods in the cyber domain by leveraging the phase-driven power behavior of GPUs and using only legitimate user-level workloads. We then establish an impedance-based analytical model that captures the interaction between GPU clusters, utility grid and the frequency-shaped dynamics of inverter-based DERs in the physical domain.

In this paper, we conduct a systematic risk analysis of these two tightly coupled infrastructures, aiming to uncover the risk of how adversaries could induce disruptions to power infrastructure through the computing infrastructure—referred to as the \texttt{Bit2Watt} risk. We further highlight that the \texttt{Bit2Watt} risk may potentially give rise to a self-reinforcing feedback loop, thereby amplifying the overall impact and triggering what we term the \texttt{Watt2Bit} risk, including denial-of-service (DoS) risks and exfiltration risks. 
The \texttt{Bit2Watt} risk exhibits a high degree of covertness, as it operates strictly within authorized workload execution paths, thereby circumventing contemporary cloud-side monitoring frameworks, creating a critical security blind spot regarding the cyber-physical ramifications induced by malicious workload orchestration. 

Our results indicate that GPU loads can reach modulation frequencies exceeding 6,000 Hz, compared with only a few hertz observed in conventional household loads such as air conditioners. Such high-frequency modulations can substantially induce voltage excursions, harmonic distortion, and damping degradation. In particular, within a 1-MW local power grid comprising 90\% DERs, manipulating just 1,000 GPUs results in a total harmonic distortion (THD) of 46.8\%, implying that nearly half of the current is expended on non-productive work and dissipated as heat—approximately 20\% more than under attack-free conditions. This not only threatens the availability of the computing equipment but also produces a negative damping ratio of -0.27, introducing an unstable mode into the system. 
Once the protections are triggered and computing loads are shed, it can trigger cascading failures, potentially leading to blackouts exceeding 80\% in large-scale power systems.

\subsection{Contributions}
First, this paper formalizes \texttt{Bit2Watt} as a new class of workload-driven cyber–physical threat, where computational processes are leveraged as \emph{controllable excitation sources} that can interact with power system dynamics. 
Unlike prior observations of workload-induced power variability, the proposed framework characterizes how such fluctuations can be systematically structured in frequency and temporal patterns, enabling persistent and targeted perturbations to power infrastructures.
This establishes a previously unexplored mechanism that links computing workload orchestration with grid-level dynamic responses, providing a principled foundation for analyzing cyber–physical risks at the interface of digital computation and physical energy systems.

\color{black}

Second, this paper develops two distinct power modulation techniques: synthetic workload modulation attack (SWMA) and LLM training modulation attack (LTMA), leveraging user-level GPU programming capabilities without requiring elevated privileges. SWMA achieves high-frequency power modulations (up to 6,000 Hz) via persistent CUDA kernels, while LTMA embeds modulation logic within standard deep learning pipelines, effectively camouflaging malicious power signatures as legitimate computational noise.

Third, this paper provides a rigorous impedance-based analysis of how GPU power modulation interacts with inverter-dominated grids. The study quantifies the impact of high-frequency GPU power modulations on power quality and system damping, offering a principled approach to understanding system vulnerabilities.

Fourth, this paper extends the impact beyond grid-side disruption by analyzing the \texttt{Watt2Bit} risk. We show that attack-induced electrical stress can create DoS conditions through protection trips and workload interruption, and further demonstrate that the same mechanism may also enable covert information exfiltration via EMI side channels. 

Fifth, this paper conducts simulation studies and real-world experiments across both the cyber computing and physical power layers. Quantitative validations are provided to illustrate the intrinsic vulnerabilities of GPU workloads and their external impact on power system stability.

\subsection{Related Work}

\textbf{Power characteristics of data centers.} 
Empirical measurements of the instantaneous power draw of an 8-GPU NVIDIA H100 HGX node were provided during the training of models, and it was shown that the power draw of the GPU fluctuated sharply during different phases \cite{IDCcharacter}.
To meet the growing power demand of modern data centers, efforts have been made to incorporate more renewable energy sources (RESs) in the power supply architecture of the data center. To cope with the intermittent nature of the RES and improve the efficiency of RES utilization, dynamic request dispatching and task scheduling of the GPUs are proposed for searching for suitable computing nodes and determining the start and finish time \cite{IDCdispatch1, IDCdispatch2, IDCDER1}, while \cite{IDCdispatch3} used the reinforcement learning based method. Techno-economic analysis and optimal sizing methodology were proposed to quantify the trade-off between power dependency and the financial cost \cite{IDCcost}.

\textbf{Interaction between grid operations and data centers.} It is found that optimally placed dispatchable computing loads, for instance, next to wind farms, can facilitate consumption of stranded power and thus enable better scaling to high renewable portfolio standard (RPS)
\cite{IDC_PG}. Microsoft, OpenAI and NVIDIA jointly proposed the challenge arising from the frequency spectrum of the power swings during different GPU tasks \cite{IDCpowersta}. They also developed innovative solutions from the perspective of the software, GPU power smoothing and rack-level energy storage. Another recent study explored the interplay between AI workload transients and data center power electronics, revealing that conventional CPU-oriented power architectures may fail to support the rapid load variations of AI accelerators \cite{GPUInflPower1}. Conversely, the power quality of the electrical grid can also affect the normal operation of GPUs in data centers. It was proved that the fluctuating power supply could influence the performance of the GPUs \cite{PowerInflGPU1}.

\textbf{Cyber and physical attacks on modern power and computing infrastructure.} Cyber–physical attacks on power infrastructure have traditionally targeted central control systems and edge devices. Prior studies show that power system operations can be disrupted through false data injection and GPS spoofing targeting control and monitoring layers \cite{targeted}. Moreover, frequency disturbances and cascading failures can be induced via large-scale botnets, including computer botnets \cite{GridShock} and appliance-based botnets \cite{madiot2.0,blackiot}. The rapid proliferation of electric vehicles (EVs) further expands the attack surface, enabling distributed load manipulation attacks that threaten grid stability \cite{EV_DLAA,EV_DLAA2}.
In computing systems, attacks primarily target the cyber domain, with side-channel attacks being a prominent example.
In particular, GPUs and heterogeneous computing platforms have been shown to leak sensitive information through shared resources and execution-dependent power signatures \cite{GPUsidechannel,timingattack,SIDECHANNEL, cacheside}.

In conclusion, existing studies have extensively analyzed GPU power characteristics and cyber-domain attacks, but largely overlook how workload-induced power variations can be exploited to impact power grid operation. This missing cross-domain perspective motivates our work, which reveals a new cyber–physical attack surface arising from the tight coupling between computing infrastructures and power systems.
\section{Background}
\subsection{Rise of Large Data Centers}
\label{sec:distributionIDC}

Data centers host large-scale CPU and GPU clusters for AI and cloud computing, requiring stable power supply. Their aggregated demand is both substantial and highly dynamic, making them not only critical loads but also potential sources of stress to the power grid.

\begin{figure} [t]
	\centering
	\includegraphics[align=t,width=0.9\linewidth,trim= 85 166 83 164, clip]{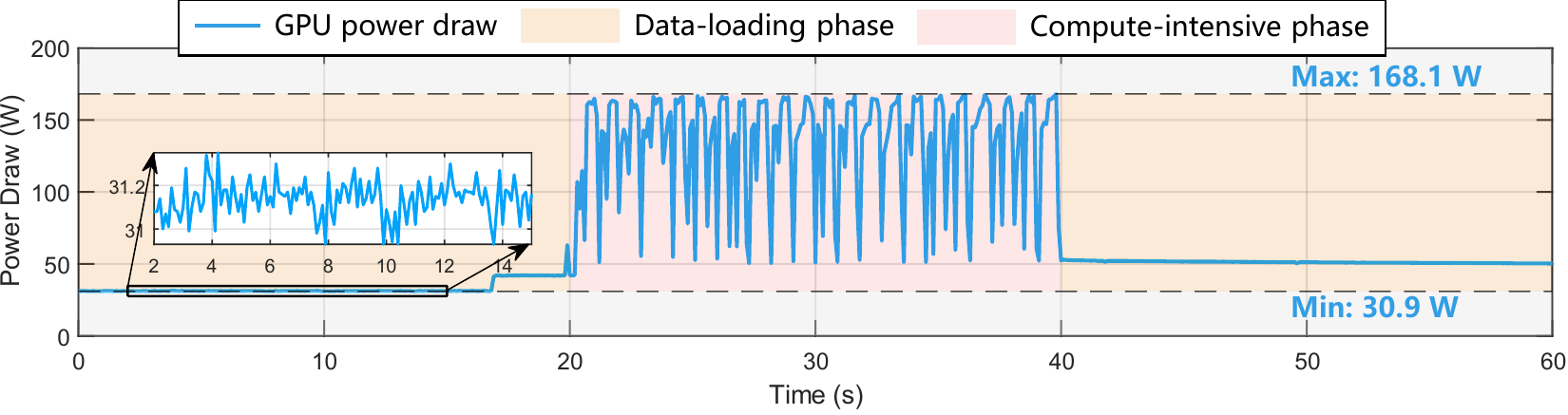}  
	\caption{Instantaneous power draw of a single NVIDIA Volta V100 GPU \cite{MITdata}.}
	\label{GPUdraw}
\end{figure}

\textbf{1. Huge electricity demand.} 
Data center deployment has rapidly expanded in recent years, with major hubs in regions such as Virginia, California, and Texas. Other spatial distribution in the U.S. is presented in the appendix. Global data center electricity consumption exceeded 400~TWh in 2024 (about 2\% of total demand) and is projected to double by 2030 \cite{IDCelectricity}. In data-center-dense regions like Virginia, this share already exceeds 25\% \cite{IDCproportion}. At the device level, as shown in Figure~\ref{GPUdraw}, during compute-intensive phases, a single V100 GPU can exhibit a power draw of nearly 170 W, approaching its hardware thermal design power (TDP). Large facilities may host over $10^5$ GPUs \cite{GPUnumber}, resulting in total power demand on the order of 20-50~MW, accounting for about one-tenth of Berlin’s total peak electrical load in 2024 \cite{BerlinPower}.

\textbf{2. Drastic power ramps.} In contrast to traditional CPU workloads that typically display smooth and predictable power consumption with lower peak-to-average ratios, GPUs demonstrate extreme power volatility. During data-loading phases, GPUs may consume only a fraction of the peak power, whereas during compute-intensive phases, the power demand can surge dramatically. The transitions between GPU workload stages often occur within milliseconds, leading to sudden spikes or drops in power demand. As illustrated in Figure~\textcolor[RGB]{50,205,50}{\ref{GPUdraw}}, the rate of change of power (RoCoP), $\Delta P / \Delta t$, is measured at nearly 1 kW/s for an individual V100 GPU and extrapolates to more than 1000 MW/s for the aggregate data center load. 

\subsection{Power Supply Architecture in Data Centers}
In recent years, the power supply architecture of data centers has undergone a profound transformation driven by the increasing integration of DERs. This shift is reflected not only in design standards such as TIA-942-C, but also in the widespread deployment of renewable-integrated power architectures in real-world data center projects. For instance, Meta Platforms has entered into multiple agreements with renewable energy developer Invenergy to procure approximately 791 MW of solar and wind power for data center operations. Similar renewable energy integration initiatives have also been undertaken by NVIDIA and other major technology companies \cite{IDCPV1, IDCPV2}.

\begin{figure} [t]
	\centering
	\includegraphics[align=t,width=1\linewidth]{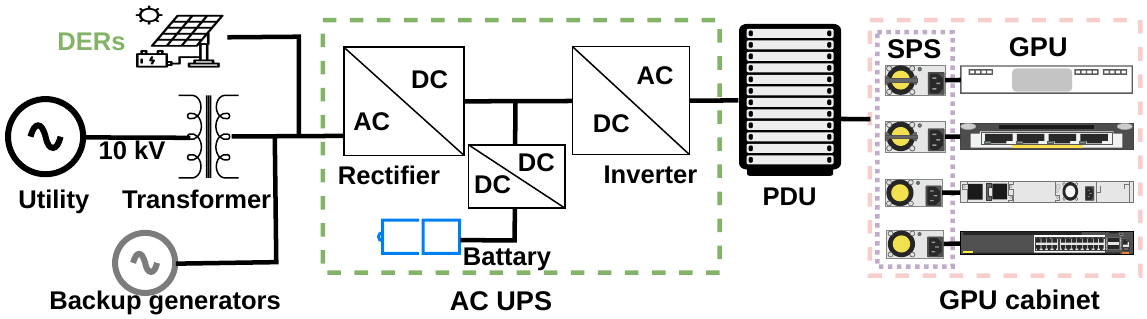}  
    \vspace{-0.1in}
	\caption{Typical power supply architecture in modern data centers.}
	\label{IDCpowerArchitecture}
\end{figure}

To guarantee high availability and fault tolerance, modern hyperscale data centers universally employ a redundant power distribution architecture (typically a 2N topology), where critical IT workloads are sustained by two independent and symmetrical power feeds. Without loss of generality, Figure \ref{IDCpowerArchitecture} presents the single-line diagram of one such power path since the redundant path shares an identical topological structure. 
As illustrated, energy from the upstream grid with high DER penetration enters the AC uninterruptible power supply (UPS) system through voltage transformation and power conversion. Subsequently, the power is routed through power distribution units (PDUs), which facilitate circuit protection and granular monitoring, before reaching the GPU cabinets. Finally, dedicated switching power supplies (SPSs) execute the terminal conversion stage to provide regulated DC power to the GPUs.

Within this power architecture, the extensive integration of the DERs, UPS, PDUs and SPSs marks a transition toward a converter-dominated system governed by sophisticated control strategies. However, this high degree of power electronization and control complexity inherently introduces significant vulnerabilities, which will be discussed later.

\subsection{Modern Power Grid}
Modern power grids are increasingly dominated by inverter-based resources (IBRs) such as PV and energy storage systems. Unlike synchronous generators, their dynamics are governed by multi-loop cascaded power-electronic control rather than physical inertia, resulting in more complex system responses. A typical PV inverter adopts a hierarchical control architecture consisting of three cascaded loops: Phase-Locked Loop (PLL) control, voltage control, and current control, as illustrated in Figure~\ref{IBRcontrolGraph}.

The PLL loop provides synchronization by estimating the grid phase angle ($\theta = \theta_0+\omega t$) and aligning the inverter with the PCC (point of common coupling) voltage $V_{pcc}$. By transforming three-phase voltages from the $abc$ frame to the rotating $dq$ frame, AC tracking is converted into a DC regulation problem. The resulting $v_d$ and $v_q$ correspond to active and reactive voltage components. Due to the trigonometric transformations, the PLL exhibits inherent nonlinearity.

The outer DC-bus voltage control loop maintains the energy balance between the input power from the PV source and the output AC power. 
The controller operates as an outer control loop that regulates the DC bus voltage by adjusting the active power exchanged with the grid. Based on the measured DC-bus voltage deviation from its reference value, the controller generates an active current reference for the inner current control loop. 

The current control loop is the innermost and fastest layer of a grid-connected DER inverter, directly governing how the inverter injects current into the grid. It ensures that the inverter output current follows the reference commands generated by the outer control loops, thereby stabilizing the inverter–grid interaction.

\section{Threat Model}

We consider a legitimate but malicious tenant operating GPU workloads in a shared computing environment (e.g., a data center connected to a DER-rich grid). The attacker requires no privileged access and instead exploits standard workload-level controls (e.g., scheduling and GPU utilization) to modulate computational intensity over time.

By coordinating high-power workloads and their start–stop timing, the attacker can induce fast and repetitive power variations at the facility level. This capability can be amplified with access to a large number of GPUs (e.g., enterprise-scale accounts).

The attacker’s objective is to inject high-frequency power fluctuations into the data center’s electrical interface. These disturbances propagate through power-electronic converters and interact with DER control loops, potentially degrading voltage/current regulations and introducing harmonics. Such effects may further feed back to the computing infrastructure (\texttt{Watt2Bit}), forming a closed-loop cyber–physical risk.

\color{black}

\section{Understanding the \texttt{Bit2Watt} Risk}
\subsection{Overview}
\begin{figure} [t]
	\centering
	\includegraphics[align=t,width=1\linewidth]{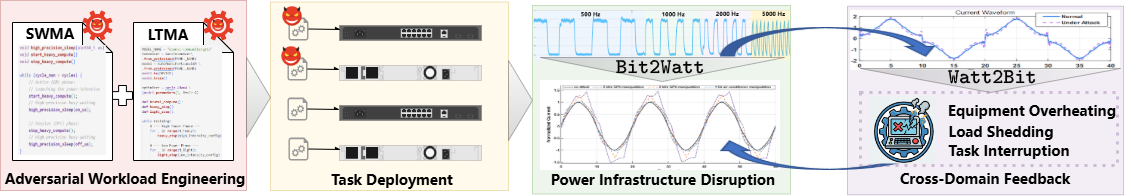}
    \vspace{-0.05in}
	\caption{Overview of the \texttt{Bit2Watt} risk. It has a four-stage cycle. It begins with the malicious workload design, i.e., malicious \texttt{Bit}. After the malicious \texttt{Bit} is deployed on the cloud platform, it induces high RoCoP and disrupts the local power grid. Power distortions in turn interrupt the computing devices via overheating and triggering the shedding protection, forming \texttt{Watt2Bit}. Finally, a vicious circle is established.}
	\label{overview}
\end{figure}

Figure \ref{overview} illustrates the four-stage execution cycle of the \texttt{Bit2Watt} vulnerability, revealing a closed-loop cyber-physical risk between computing workloads and power infrastructure.
To understand the \texttt{Bit2Watt} risk, we first elaborate on: 1) how to induce power fluctuation through GPU tasks and 2) how the power fluctuation affects the power infrastructure, and we validate these principles with simulations and real-world experiments.

\subsection{How to induce power fluctuation through GPU tasks?}

\subsubsection{Principle and Design}
The instantaneous GPU power draw is tightly coupled with the computational workload running on streaming multiprocessors (SMs), tensor cores, memory controllers, and the associated on-chip clock generators. When computational intensity increases, the GPU automatically raises frequency and current to deliver higher performance, resulting in a rapid increase in dynamic power consumption. Conversely, when workload demand drops, the GPU transitions toward energy-saving states.

This behavior provides a software-accessible mechanism for power modulation without requiring privileged access. By controlling GPU workloads, an attacker can induce oscillations between high- and low-power phases. In this subsection, we design two attack methods based solely on user-level GPU programming capabilities.

\textbf{1. Synthetic Workload Modulation Attack (SWMA).} The attacker uploads a purpose-built CUDA program consisting of two internal execution modes:
1) active mode, which executes high-intensity tensor-core or FP32 FMA loops with large global-memory access, saturating SM occupancy and maximizing power consumption;
2) passive mode, in which the kernel performs negligible computation and quickly enters an idle state.

The implementation of SWMA relies on a GPU kernel and a CPU controller. The GPU kernel features two internal execution modes, while the CPU controller communicates with the kernel via a unified-memory control flag allocated using \texttt{cudaMallocManaged}, enabling real-time modulation of the kernel's execution mode.
Once the kernel is launched, it enters a continuous polling loop in which each thread block monitors the control flag and instantly transitions to the corresponding execution mode upon detecting a change. Meanwhile, a CPU-side controller periodically updates the flag according to a predefined switching schedule. The controller itself runs entirely in user space and therefore requires no privileged access to GPU hardware or system firmware.

\textbf{2. LLM Training Modulation Attack (LTMA).} This approach exploits the intrinsic compute variability within Large Language Model (LLM) training to induce power modulations without designing an additional synthetic workload. Instead of launching a GPU kernel, the adversary embeds modulation logic directly into the training pipeline of an LLM. Modern distributed training frameworks (e.g., PyTorch, DeepSpeed, Megatron-LM) execute complex sequences of compute-heavy primitives—such as tensor-core GEMM kernels, fused attention kernels, and backpropagation primitives—whose instantaneous intensity and invocation patterns strongly influence GPU power draw.

In this design, the training script functions as the modulation controller. The control is achieved by dynamically manipulating hyperparameters or inserting auxiliary operations at specific time intervals. During the high-load phases, the training script intentionally increases instantaneous compute demand through legitimate mechanisms. The low‑load phase can be produced through modifications that temporarily lighten the computations without compromising training correctness.

This attack requires no elevated privileges: cloud tenants ordinarily possess full control over their training scripts, batch schedules, and dataflow definitions. Therefore, the modulation is implemented entirely in user-space within the training framework, blending seamlessly into normal training behavior and leaving no externally visible synthetic kernels. 

\textbf{Comparison.} The SWMA could offer precise control over frequency and duty cycle through the host-side modulation logic, and support higher-frequency switching. However, its reliance on a custom CUDA kernel and device-level polling might reduce the concealment. LTMA operates entirely within standard Python-level training pipelines, making it far easier to implement and far more difficult for the cloud provider to distinguish from legitimate training noise. Its control accuracy is inherently constrained by the iteration rate of the training loop. 

\subsubsection{Feasibility Analysis}
\textbf{Availability.} Modern hyperscale cloud platforms, including AWS, Google Cloud, Alibaba Cloud, Huawei Cloud, and Azure, provide tenants with on-demand, elastic access to powerful HPC infrastructure. This pervasive availability of rentable GPU-accelerated compute has dramatically lowered the barrier for running large-scale AI and data-intensive workloads.
These corporations expose unified SDKs, REST APIs, and CLI tools that enable tenants to automate job submissions, upload containerized training tasks, and precisely control start–stop timing.

\textbf{Spatial aggregation.} Existing schedulers use topology-aware placement, which strategically maps distributed tasks—particularly large-scale AI training—to physically adjacent GPU nodes within the same rack to exploit high-bandwidth interconnects like NVLink and InfiniBand. This feature inadvertently facilitates the physical aggregation of electrical demand, thereby avoiding the geographical dispersion of power fluctuations from different GPUs, which would otherwise dilute their aggregate impact.

\textbf{Temporal synchronization.} 
Adversaries could utilize user-space busy-waiting loops on high-resolution timestamps, thereby reducing the non-deterministic scheduling jitter of the operating system. In HPC or dedicated GPU clusters, this synchronization can be achieved via Precision Time Protocol (PTP), whereas in public clouds it is typically provided implicitly by provider-managed clock services. Nevertheless, achieving perfectly aligned power transitions between high-power and low-power phases across all GPUs is still a challenge.

\textbf{Cost.} An adversary can readily acquire legitimate access to these resources by obtaining valid tenant accounts or credentials, operating under the guise of a normal user. The competitive pricing of high-performance GPU rental—ranging from \$0.39/hour to \$2.99/hour \cite{GPUprice1, GPUprice2}—further lowers the financial threshold for mounting an attack. Although these costs are typically quoted per hour, a synchronized attack often requires an execution time of only a few minutes to degrade the power quality and stability, significantly reducing the actual cost and time commitment for the adversary.

\subsubsection{Validation}

\textbf{Experiment setup.} 
The data were captured on modified in-lab workstations running Docker, equipped with INTEL(R) XEON(R) SILVER 4510 processor and 128~GB of SK Hynix HMCG78AEBRA107N DDR5 RDIMM memory, configured as eight 16~GB DIMMs operating at 4400~MT/s.
The LTMA is realized through the training process of GPT-2. Power data are measured using an NI USB-6421 data acquisition system (DAQ) in conjunction with a LabVIEW-based data acquisition application. 
The DAQ senses the current and voltage signals from the GPU-side auxiliary +12~V supply lines.

\begin{table}[t]
\centering
\caption{Power modulation results of representative GPUs}
\label{tab:gpu_power_modulation}
\small
\begin{tabular}{@{}lccc cc@{\hspace{10pt}}c@{\hspace{10pt}} cc@{\hspace{10pt}}c@{}}
\toprule
\multirow{3}{*}{\textbf{GPU}} &
\multirow{3}{*}{\textbf{Arch.}} &
\multirow{3}{*}{\textbf{Year}} &
\multirow{3}{*}{\textbf{TDP}} &
\multicolumn{3}{c}{\textbf{SWMA}} &
\multicolumn{3}{c}{\textbf{LTMA}} \\
\cmidrule(lr){5-7} \cmidrule(lr){8-10}
 & & & 
 & \textbf{Freq.} & \textbf{Amp.} & \textbf{Ctrl.}
 & \textbf{Freq.} & \textbf{Amp.} & \textbf{Ctrl.} \\
 & & & \textbf{(W)}
 & \textbf{(Hz)} & \textbf{(W)} & 
 & \textbf{(Hz)} & \textbf{(W)} & \\
\midrule
RTX 2080 & Turing & 2018 & 185 & 1{,}500 & 103 & \checkmark & 1{,}200 & 134 & $\times$ \\
RTX 3080 & Ampere & 2020 & 320 & 2{,}000 & 157 & \checkmark & 1{,}800 & 225 & $\times$ \\
RTX 3090 & Ampere & 2020 & 350 & 5{,}000 & 204 & \checkmark & 2{,}000 & 265 & $\times$ \\
RTX 4090 & Ada Lovelace & 2022 & 450 & 6{,}000 & 248 & \checkmark & 3{,}000 & 355 & $\times$ \\
Tesla V100 & Volta & 2017 & 250 & 3{,}000 & 124 & \checkmark & 2{,}000 & 210 & $\times$ \\
A100 & Ampere & 2020 & 250 & 5{,}000 & 135 & \checkmark & 3{,}000 & 215 & $\times$ \\
\bottomrule
\end{tabular}
\begin{tablenotes}
	\item $\dagger$ Arch.: architecture; \ \  Freq.: frequency;\ \  Amp.: amplitude; \ \  Ctrl.: controllability.
\end{tablenotes} 
\end{table}

\begin{figure} [t]
	\centering
	\includegraphics[align=t,width=1\linewidth]{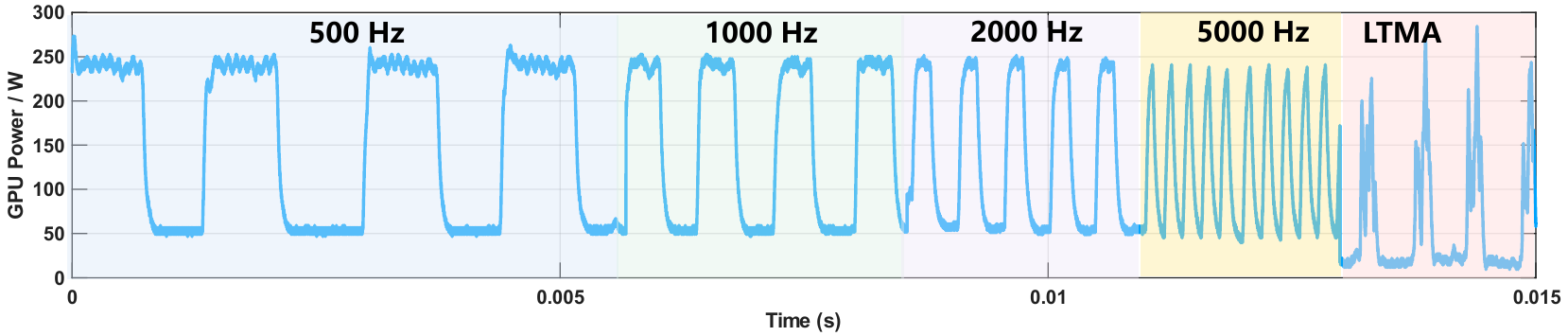}  
	\caption{Concatenated results of different power modulation methods on RTX 3090. The first four attacks are SWMA and the last is a 2000 Hz LTMA.}
	\label{GPUpower}
\end{figure}

\textbf{Result analysis.} Table~\ref{tab:gpu_power_modulation} summarizes the power modulation capability of representative NVIDIA GPUs across both consumer-grade and data-center architectures. 
Across all tested GPUs, SWMA achieves higher modulation frequencies than LTMA, with spectral components ranging from 1.5~kHz to 6~kHz. This is enabled by fine-grained workload scheduling and kernel-level execution dynamics. In contrast, LTMA exhibits lower modulation frequencies (1.2–3~kHz) but with larger modulation amplitudes. The time-series power profiles under different modulation strategies are concatenated and illustrated in Figure \ref{GPUpower}.
Taking the 5,000~Hz SWMA as an example, when the modulation frequency exceeds this level, the fluctuation amplitude can no longer reach 204~W. This limitation is attributed to the finite dynamic response of GPU power management and voltage regulation, which cannot fully track rapid workload transitions.

From a controllability perspective, SWMA allows explicit and precise tuning of the modulation frequency via workload scheduling, whereas LTMA lacks such direct control. In LTMA, both frequency and amplitude are implicitly determined by training dynamics (e.g., batch processing, optimizer updates, and synchronization), making them task-dependent and less predictable.

The observed extreme modulation frequency of up to 6,000 Hz suggests that such loads have the potential to excite high-frequency resonance modes in DER systems, which are usually not considered in conventional stability analyses.

\begin{figure} [t]
	\centering
	\includegraphics[align=t,width=1\linewidth]{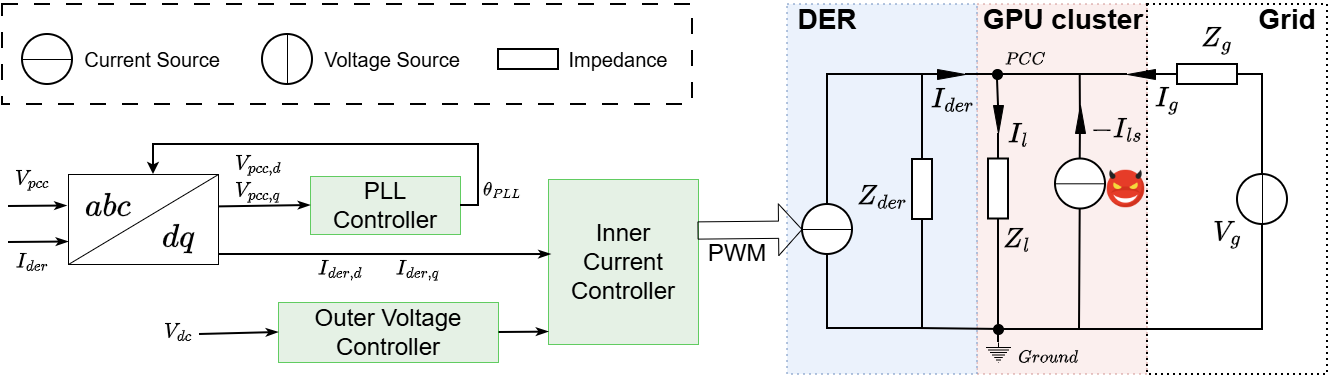}  
	\caption{Typical power supply system of the local data center with integration of DERs.}
	\label{IBRcontrolGraph}
\end{figure}

\subsection{How the power fluctuation affects the power infrastructure?}

\subsubsection{Fundamental Model}

To capture the interaction mechanism, we adopt an impedance-based model using Thévenin and Norton equivalents for the cascaded power supply system \cite{impedance0}. The system consists of three components connected at the PCC: the main grid, the converter-based DERs, and the GPU cluster, as shown in Fig.~\ref{IBRcontrolGraph}.

The main grid is represented by a Thévenin equivalent, i.e., an ideal voltage source $V_g$ with an equivalent impedance $Z_g$. The DER is modeled as a controlled current source with an output impedance $Z_{der}$ shaped by its internal control loops.
The GPU cluster behaves as a near constant-power load (CPL) at the PCC due to multi-stage power-electronic regulation \cite{GPUCPL}. 
CPL represents an intrinsic short-term electrical behavior, distinct from longer-term power fluctuation.
Therefore, the GPU cluster is modeled as a workload-controlled current sink $-I_{ls}$ in parallel with a CPL impedance $Z_{l}$, where the negative sign of the controlled current indicates power absorption from the PCC rather than current injection into the grid.

In the subsequent subsections, we first clarify the mechanism by which GPUs respond to workload-dependent power demand, where workload-induced power fluctuations are manifested as current fluctuations and interact with the DER and grid impedance. 
We then analyze the internal vulnerabilities that make the system sensitive to such fluctuations. 
Finally, we explain the potential impacts caused by these fluctuations.

\subsubsection{Power Adjustment Mechanism of GPU Clusters}
\label{sec:powermechanism}
At the board level, the single GPU's internal voltage regulator module (VRM) maintains a constant core voltage ($V_{core}$) within its control bandwidth (typically tens of kHz), forcing the power demand to be supplied primarily through rapid current adjustments.
At the system level, upstream power conditioning stages regulate the intermediate DC bus voltage by modulating the AC-side current. This energy-balancing mechanism propagates the GPU's power variations to the grid interface in the form of current fluctuations.

\subsubsection{Internal Vulnerability Analysis}
\textbf{1. The negative incremental resistance of the GPU clusters:} The defining characteristic of the CPL is its negative incremental resistance (NIR), as small changes in voltage ($\Delta v$) result in inverse changes in current ($\Delta i$) to maintain constant power ($P = V \cdot I$). 
The impedance of the load $Z_{l}$ in Figure \ref{IBRcontrolGraph} is approximated by a negative real part:
$Z_{l}(s) \approx -V_{l}^2/P_{l}$.
NIR could amplify disturbances through a positive feedback mechanism. With grid impedance $Z_g$, the PCC voltage is $V_{pcc} = V_g - I_g \cdot Z_g$. If $V_{pcc}$ drops due to the increase of $I_{gpu}=I_l+I_{ls}$, the CPL further increases $I$ to maintain $P$, which further decreases the voltage across $Z_g$, worsening the fluctuation, and vice versa. Meanwhile, the negative resistance could reduce the system damping, potentially degrading stability margins.

\textbf{2. Control-induced frequency response of the DER output impedance:}
Unlike synchronous generators, DERs have negligible physical inertia, so their voltage and frequency responses rely mainly on control dynamics. 
It results in a complex characteristic frequency response $Z_{der}(s)$ and exhibits resonant behavior, where disturbances at certain frequencies can be amplified, leading to voltage and current distortions and oscillations.

Although modern systems employ control and protection mechanisms like virtual inertia emulation and over-current/voltage protection schemes, they are often limited by practical constraints, such as communication delays, measurement errors, parameter uncertainty, and insufficient coordination under high DER penetration. 
High-frequency fluctuations induced by GPU clusters can exceed their response bandwidth, making these mechanisms ineffective and exposing system vulnerabilities.

\subsubsection{Validation}
In this subsection, we validate the above mentioned internal vulnerabilities, respectively.

\begin{figure} [t]
\centering
\subfigure[Experiment setup]
{
        \label{CPLsetup}
		\includegraphics[align=t,width=0.48\linewidth]{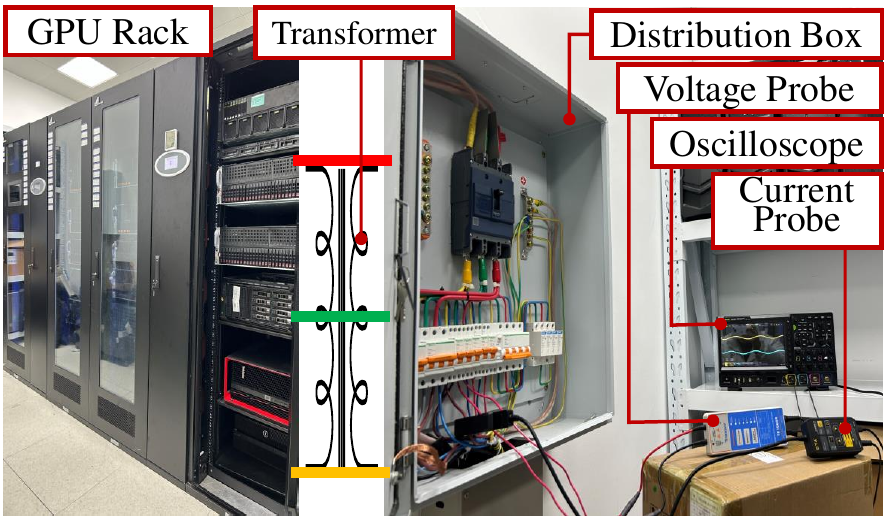}} 
\subfigure[Electrical characteristic of GPU clusters]
{
        \label{CPL}
		\includegraphics[align=t,width=0.44\linewidth]{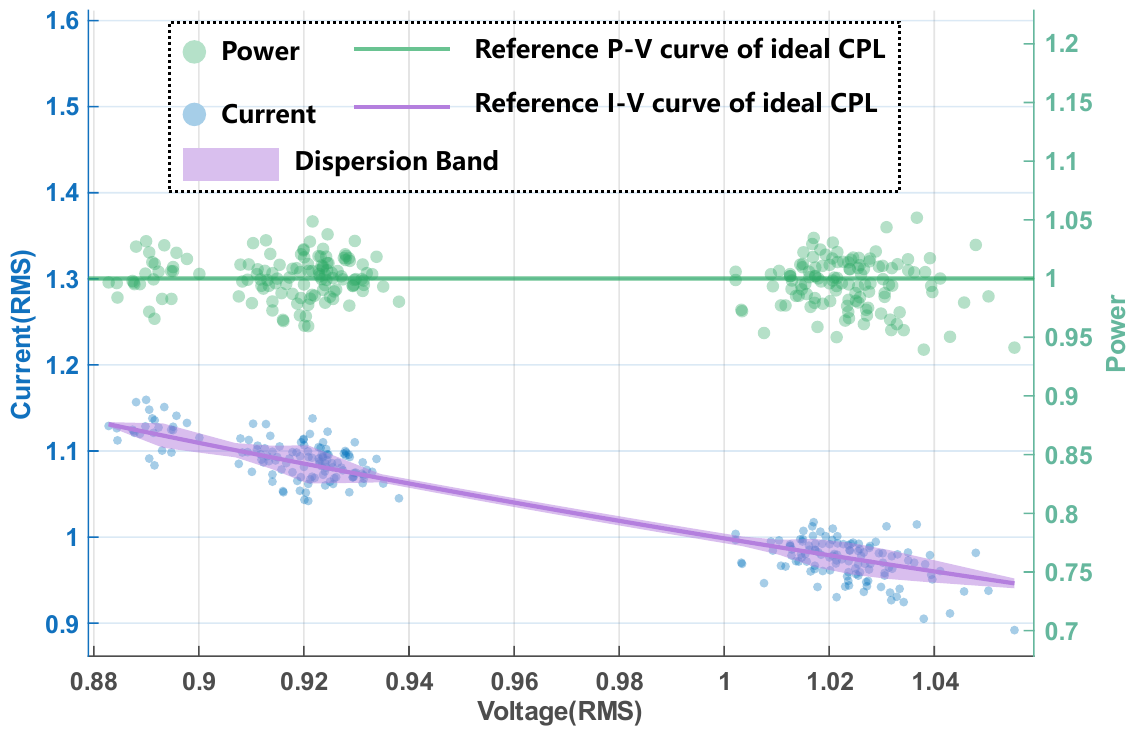}}        
        \caption{Validation of the CPL and NIR of the GPU clusters.}
	\label{CPL_experiment}
\end{figure}

\textbf{1. Validation of the CPL and NIR of the GPU clusters:} We conduct real-world voltage and current measurement experiments on a GPU cluster testbed shown in Fig. \ref{CPLsetup}. Abrupt voltage perturbations are intentionally introduced by adjusting the transformer, and the corresponding root-mean-square (RMS) voltage and current are simultaneously measured at the electrical interface of the platform.

\textbf{Experiment setup.} The tested GPU cluster is hosted in a dedicated server room with standard industrial cooling infrastructure and is powered by a three-phase AC UPS (Envicool ER010L). The cluster comprises three GPU racks with approximately 100 GPUs, including NVIDIA A100, V100, and RTX-series. Power for the entire room is supplied through the distribution box in Figure \ref{CPLsetup}. Therefore, measurements taken at the distribution box capture the collective electrical behavior of the GPU cluster and serve as a representative interface for this small-scale deployment. 
The measuring tools include the current probe ETA5301M, the voltage probe TR1005B and the oscilloscope RIGOL MSO8064. Signals are sampled at a rate of 10 MSa/s.

\textbf{Result analysis.} To mitigate incidental effects and improve statistical robustness, we applied abrupt voltage perturbations in ten independent trials, including five voltage sag events (approximately 225 V $\rightarrow$ 195 V) and five voltage swell events (approximately 195 V $\rightarrow$ 225 V). The normalized statistical $I$-$V$ and $P$-$V$ trajectory results are depicted in Figure \ref{CPL}.

As illustrated, the RMS current exhibits an inverse dependence on the RMS voltage, with measured samples closely distributed around the theoretical CPL reference curve ($I \propto 1/V$). The power consumption data (green scatter points) remain tightly concentrated around the nominal baseline (green reference line). The mean error of the power measurements is only $0.0151$, with a standard deviation of $\sigma = 0.0123$. In other words, voltage steps exceeding 10\% induce power variations of merely 1\%, providing empirical evidence of short-term constant-power behavior and NIR characteristics.

The slight dispersion and occasional outliers observed in the data points can be attributed to the composite nature of the data center. In addition to the primary GPU computational loads, the system inevitably incorporates non-CPL auxiliary components, such as resistive lighting elements and frequency-controlled cooling fans. These loads introduce a minor constant-impedance behavior, accounting for the measured variances. Specifically, points in the shaded dispersion band ($\pm1\sigma$) reflect the inherent power variability.

\textbf{2. Validation of the frequency response of the DER output impedance:} We scan the impedance frequency response using the perturbation injection method, by injecting small-amplitude sinusoidal current perturbations at selected frequencies and extracting the corresponding output voltage components via FFT-based spectral analysis.

\begin{figure} [t]
\centering
\subfigure[Experiment setup]
{
        \label{SetupImpedance}
		\includegraphics[align=t,width=0.45\linewidth]{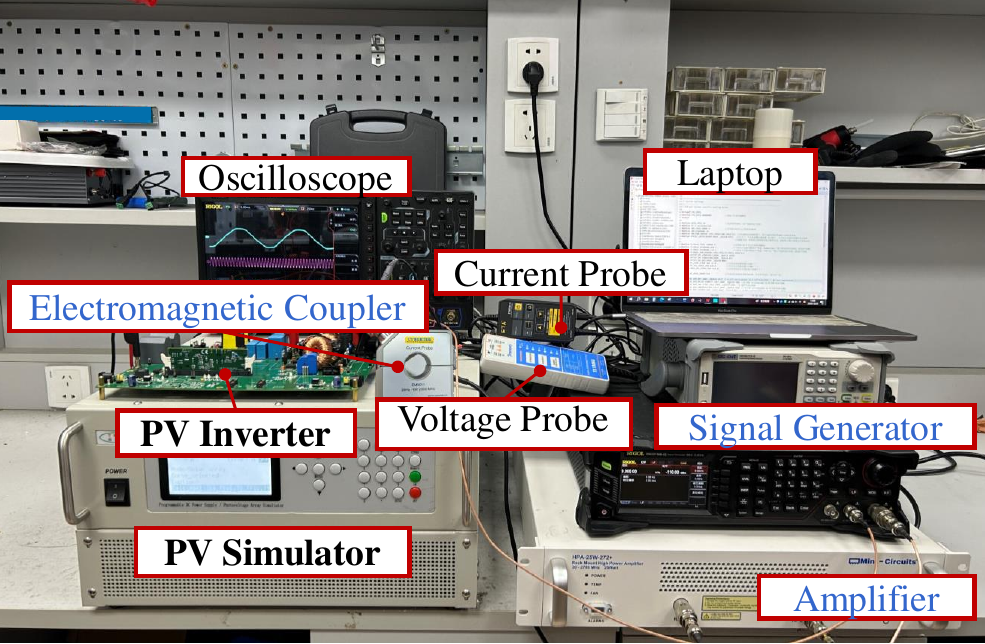}} 
\subfigure[Bode plots of grid and DER impedance]
{
        \label{impedance_interaction}
		\includegraphics[align=t,width=0.465\linewidth]{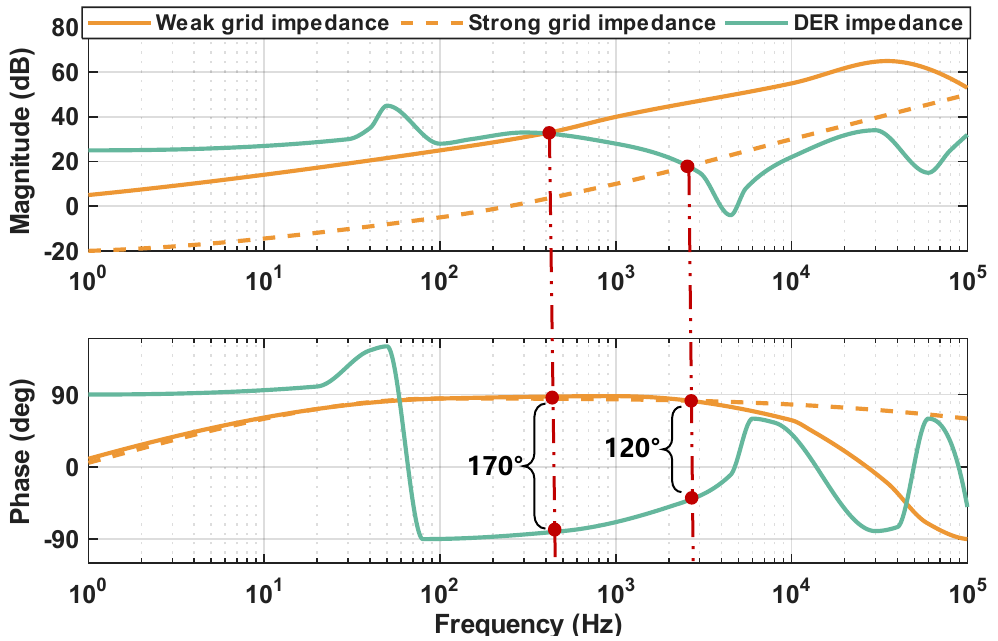}}        
\caption{Validation of the frequency response of the DER output impedance.}
\label{impedance_experiment}
\end{figure}

\textbf{Experiment setup.} As shown in Figure \ref{SetupImpedance}, the tested devices include the inverter (TI C2000 inverter development kit) \cite{TIinverter}, and a programmable solar panel emulator (TEWERDTPV1000) to emulate its upstream solar panels.
We use the SIGLENT SDG6032X-E signal generator to generate signals from 1 Hz to 10 kHz, use amplifier HPA-25W-272+ to amplify it to 25 W, and couple the signal into the circuit with the electromagnetic coupler EM5011.
The data are acquired with the current probe ETA5301M, voltage probe TR1005B and the oscilloscope RIGOL MSO8064.

\textbf{Result analysis.} 
As shown in Fig.~\ref{impedance_interaction}, the DER output impedance exhibits a frequency-dependent profile with resonant peaks at specific high frequencies, indicating increased sensitivity to high-frequency perturbations, i.e., ``out-of-band'' vulnerabilities. As a result, GPU-induced current fluctuations may amplify harmonic components, leading to power quality degradation and oscillations.

Fig.~\ref{impedance_interaction} also illustrates the interaction between the DER and grid impedance. According to the Middlebrook criterion, system stability depends on the ratio between grid impedance $Z_g$ and the equivalent load impedance $Z_{dl} = Z_l \parallel Z_{der}$, with stability maintained when
\begin{equation}
\left|T_m(j \omega)\right| = \left| \frac{Z_g(j\omega)}{Z_{dl}(j\omega)} \right| < 1.
\end{equation}
The phase margin is defined as $PM = 180^\circ-\left|\angle T_m(j \omega_c)\right|$, where $\omega_c$ is the crossover frequency that satisfies $|T_m| = 1$, i.e., $|Z_g(j\omega_c)| = |Z_{dl}(j\omega_c)|$. 
As DER penetration increases, the system transitions toward a weak grid, significantly reducing the stability margin (e.g., from $\sim60^\circ$ to $\sim10^\circ$). The negative incremental resistance of GPU clusters further reduces the remaining margin, exacerbating instability.

\subsubsection{Impact Analysis of the External Fluctuation}

\textbf{1. Power quality.} 
The impact on power quality includes voltage excursions and harmonics. Due to the extremely fast power ramps of GPU clusters, the current at the PCC must change rapidly to maintain power balance. In regulated systems where voltage variations are tightly constrained, these fast current transients propagate through the network impedance $Z_{g}$ and result in noticeable voltage disturbances: $\Delta V(t) = Z_g \cdot \Delta I(t).$

Beyond voltage excursions, the high-frequency current variations $dI/dt$ induced by GPU power ramps generate harmonic components that propagate through the network and degrade power quality. When these components align with resonant frequencies of DER control loops or grid elements (e.g., LCL filters), they can be significantly amplified.

\textbf{2. System instability.} 
In practical conditions, there are plenty of nonlinear elements that cannot be ignored when the system suffers from attacks, such as amplitude limiting, time delay, dead zone, saturation and complex power electronic switching and control (e.g., PLL) in DERs.
The high-frequency power modulations from GPU clusters fundamentally alter the system’s nonlinear operating regime.
Let the nonlinear system be represented as $\bm{\dot x}=\bm{f(x,\mu)}$, where $\bm{x}$ denotes the system state vector, and $\mu$ is the disturbance (in this case it is the power fluctuation of the GPU clusters).

Typically, it operates around an equilibrium $\bm{x}_0$, whose stability is determined by the Jacobian matrix $\bm{J}=\frac{\bm{\partial f}}{\bm{\partial x}} |_ {\bm{x_0, \mu_0}}$. Under the GPU-induced power modulations, the effective operating point of the system shifts, modifying the Jacobian matrix $\bm{J}(\mu)$ and dominant eigenvalues $\lambda_{1,2}(\mu)=\alpha \pm j\beta(\mu)$. A critical transition is triggered when $\mu$ reaches a threshold $\mu_c$ satisfying the Hopf bifurcation conditions: $\alpha(\mu_c) = 0,\ \ \beta(\mu_c) = \omega_{osc} \neq 0$ and $\frac{d\alpha}{d\mu} |_{\mu_c} > 0$.
Physically, this reflects the reduction of effective damping due to the interaction between GPU-induced load dynamics and DER control loops.

\subsection{\texttt{Watt2Bit} Risk}
The above analysis focuses on the impact of GPU power manipulation on power systems. A critical yet overlooked question arises: \emph{can the induced power disturbances, in turn, affect the computing domain?} The answer is yes and we call it the \texttt{Watt2Bit} risk. 

\textbf{DoS.}
The cyber–physical interaction does not terminate at the power system boundary, but forms a closed-loop cyber–physical vulnerability.
Due to the nonlinear switching mode of the power supply in data centers, the mechanisms behind the \texttt{Watt2Bit} can be complex and various; here we illustrate it via harmonic distortions. 
High-frequency harmonics can lead to elevated resistive losses and increased thermal stress within power conversion stages due to increased RMS current and the frequency-dependent rise in equivalent series resistance (ESR). This may trigger protection mechanisms (e.g., over-temperature or over-current), interrupting GPU operation and causing denial of service (DoS). Consequently, watt-level electrical disturbances propagate into bit-level computational disruptions.

\textbf{Exfiltration.} The attack may also act as an side-channel exfiltration vector. In particular, an adversary could encode bits by modulating either the attack frequency or the attack amplitude, thereby inducing distinguishable patterns in the measured power/EMI traces. A receiver observing these side-channel emissions could then demodulate the corresponding spectral or amplitude features to recover the transmitted symbols. 

\color{black}

\begin{table}[t]
\caption{Electrical parameters of the Simulink model.}
\label{tab:impedance}
\centering
\footnotesize
\setlength{\tabcolsep}{3pt}
\renewcommand{\arraystretch}{1.08}
\begin{tabularx}{\columnwidth}{@{}l X@{}}
\toprule
Block / stage & Value / description \\
\midrule
PV array &
$N_{\mathrm{series}}=100$, $N_{\mathrm{parallel}}=47$; per-module $P_m=214.97$~W. \\

PV-side boost filter &
Input shunt capacitor $C=4.7$~mF, ESR $=1$~m$\Omega$; boost inductor $L=20$~mH, $R=1$~m$\Omega$; output RC branch $C=4.7$~mF, ESR $=1$~m$\Omega$. \\

Grid source &
$400$~V, $50$~Hz; source resistance $=10^{-5}~\Omega$; short-circuit level $=100$~MVA. \\

Feeder line &
Length $=5$~km; $r=[0.01273,\ 0.3864]$ $\Omega$, $l=[0.9337,\ 4.1264]$~mH, $c=[12,\ 12]$~nF. \\

Bridges &
$R_{\mathrm{on}}=1$~m$\Omega$; snubber resistance $=100$~k$\Omega$; snubber capacitance $=\infty$. \\

UPS output filter &
Per-phase RL filter with $R=0.05~\Omega$ and $L=2$~mH. \\

\bottomrule
\end{tabularx}
\end{table}

\begin{table}[t]
\caption{DER, UPS and SPS control settings in the Simulink model.}
\label{tab:control_settings}
\centering
\footnotesize
\setlength{\tabcolsep}{3pt}
\renewcommand{\arraystretch}{1.08}
\begin{tabularx}{\columnwidth}{@{}l X@{}}
\toprule
Loop / block & Value / description \\
\midrule

PV DC/DC duty control &
\texttt{PI Controller} with $K_p=0.001$, $K_i=0.01$; $f_s = 5$~kHz. \\

PLL PI regulator &
$K_p=10$, $K_i=5\times 10^4$; $q$-axis voltage error as the locking signal \\

PV inverter $dq$ outer loop &
Voltage \texttt{PI Controller}: $K_p=0.1$, $K_i=100$; $V_{PV,d}^\ast=285$~V, $V_{PV,q}^\ast=0$~V. \\

PV inverter $dq$ inner loop &
Current \texttt{PI Controller} with $K_p=30$, $K_i=200$. \\

UPS rectifier outer loop &
DC-bus voltage \texttt{PI Controller} with $K_p=5$, $K_i=0.001$; $V_{ref}^\ast=500$~V. \\

UPS rectifier inner $dq$ loop &
Current \texttt{PI Controller} with $K_p=1$, $K_i=0.1$; reference $i_q^\ast=0$. \\

UPS DC/DC loop &
DC-bus voltage \texttt{PI Controller} with $K_p=0.002$, $K_i=0.05$; reference $V_{\mathrm{rectiDC}}^\ast=850$~V; $f_s = 20$~kHz. \\

UPS inverter outer loop &
Voltage \texttt{PI Controller} with $K_p=0.01$, $K_i=12$; reference $V_Q^\ast=0$. \\

UPS inverter inner loop &
Current \texttt{PI Controller} with $K_p=1.0$, $K_i=120$. \\

SPS rectifier outer loop &
DC-bus voltage \texttt{PI Controller} with $K_p=0.18$, $K_i=6.0$; $V_{ref}^\ast=500$~V. \\

SPS rectifier inner loop &
Current \texttt{PI Controller} with $K_p=1.2$, $K_i=20.0$; reference $i_q^\ast=0$. \\

\bottomrule
\end{tabularx}
\end{table}

\begin{figure} [h]
\centering
\subfigure[Simulation topology]
{
        \label{SimTopo}
		\includegraphics[align=t,width=0.56\linewidth]{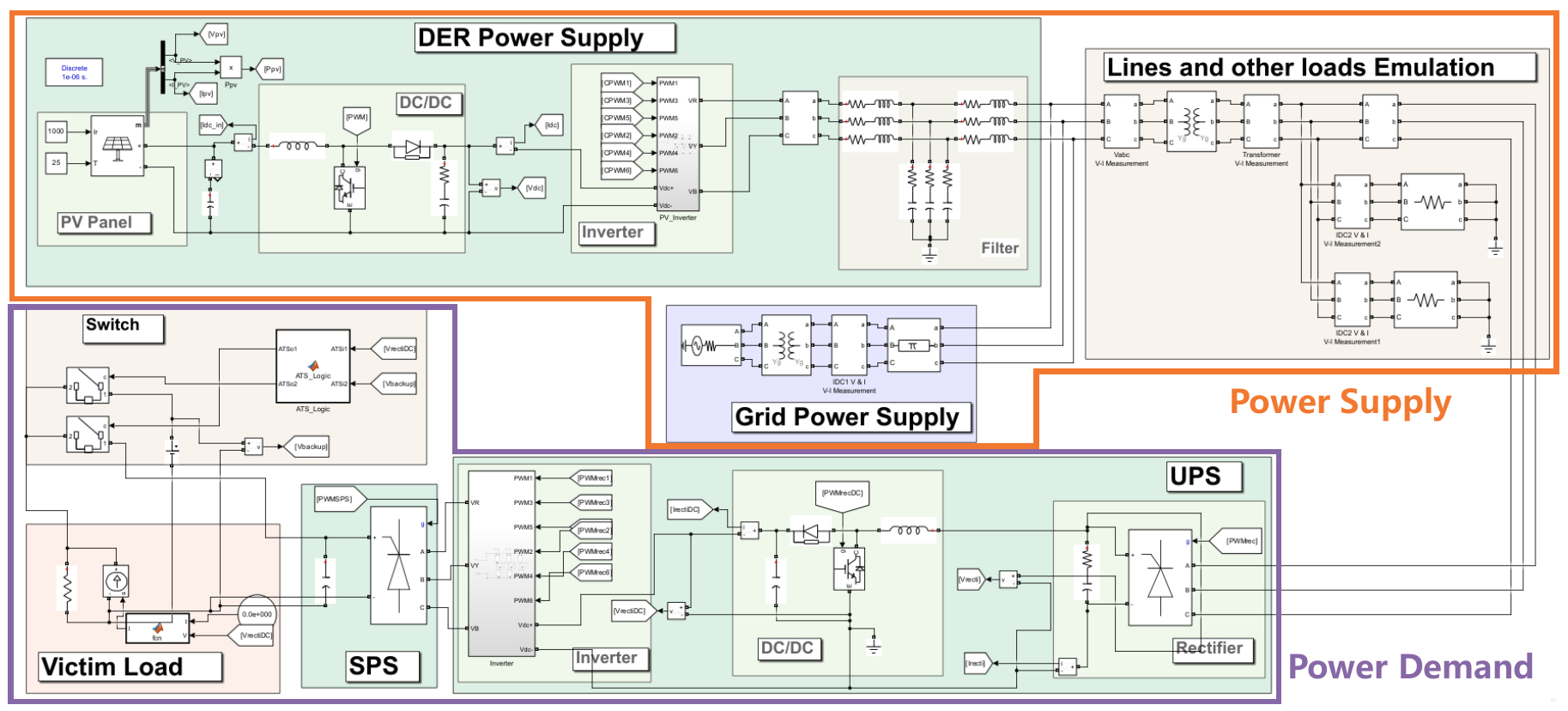}} 
\hspace{-0.9em} 
\subfigure[IEEE 118-bus system ]
{
        \label{118bus}
		\includegraphics[align=t,width=0.42\linewidth]{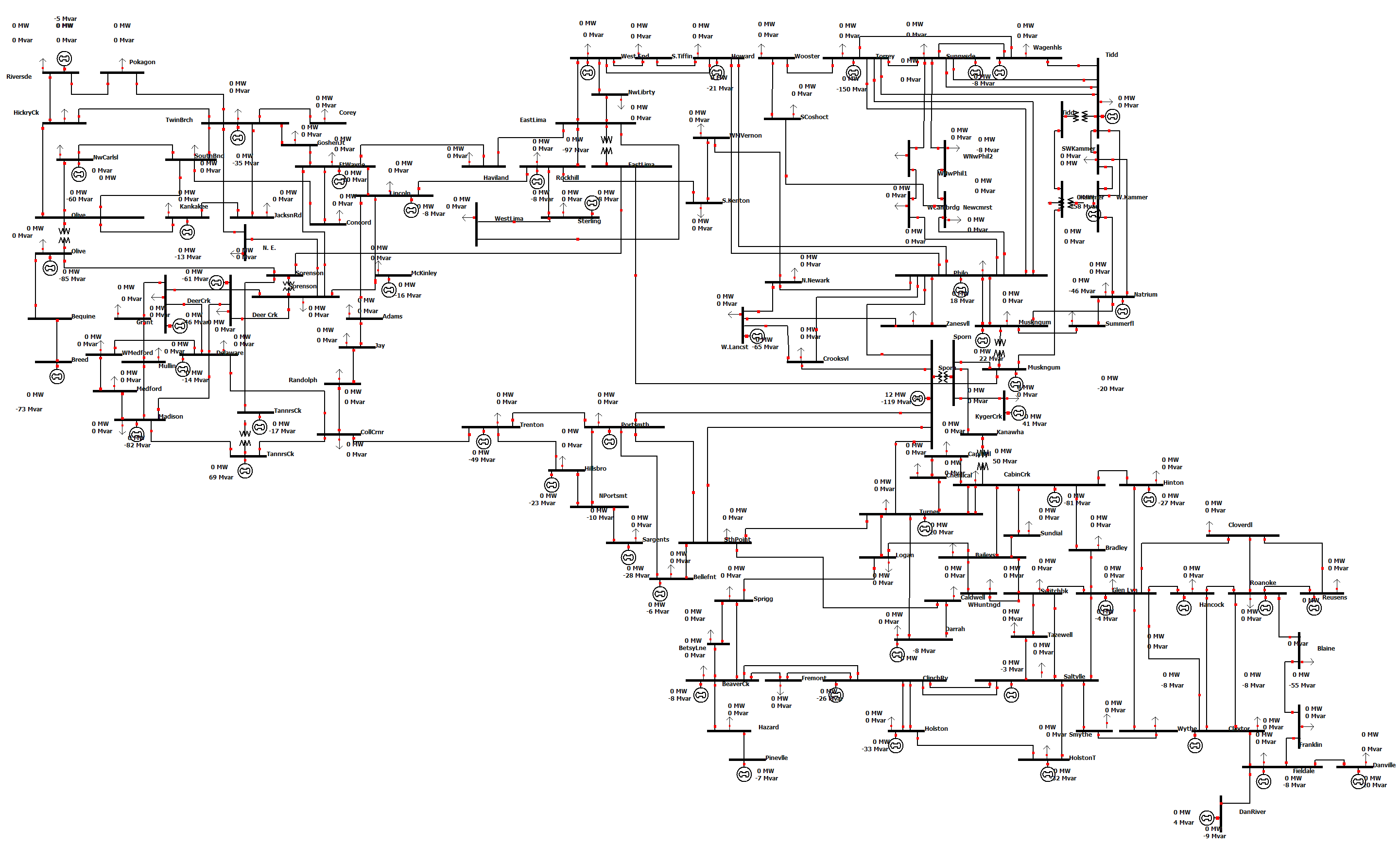}}
\caption{System topologies used in the simulation.}
	\label{SimulationTopo}
\end{figure}

\begin{figure} [t]
\centering
\subfigure[Voltage excursion]
{
        \label{V_impact}
		\includegraphics[align=t,width=0.33\linewidth]{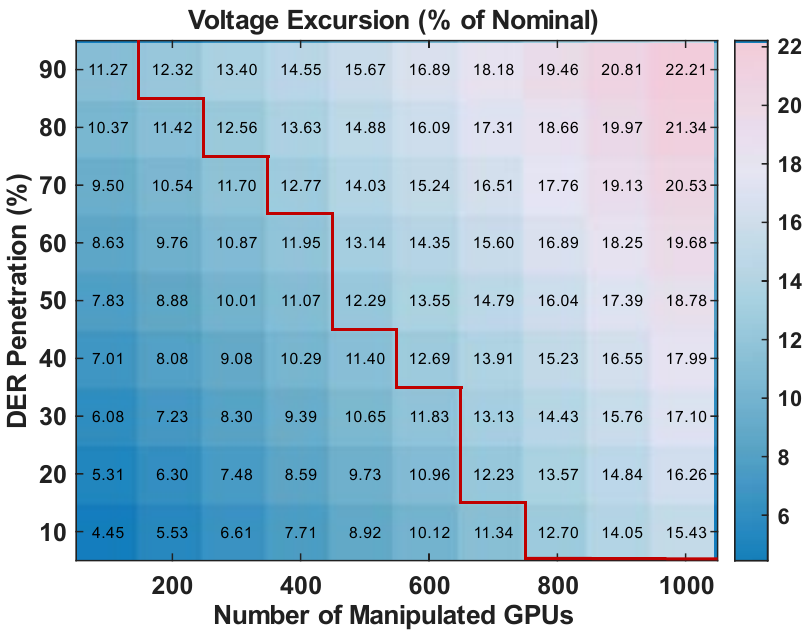}} 
\hspace{-0.9em} 
\subfigure[Harmonics (Current THD)]
{
        \label{THD_impact}
		\includegraphics[align=t,width=0.33\linewidth]{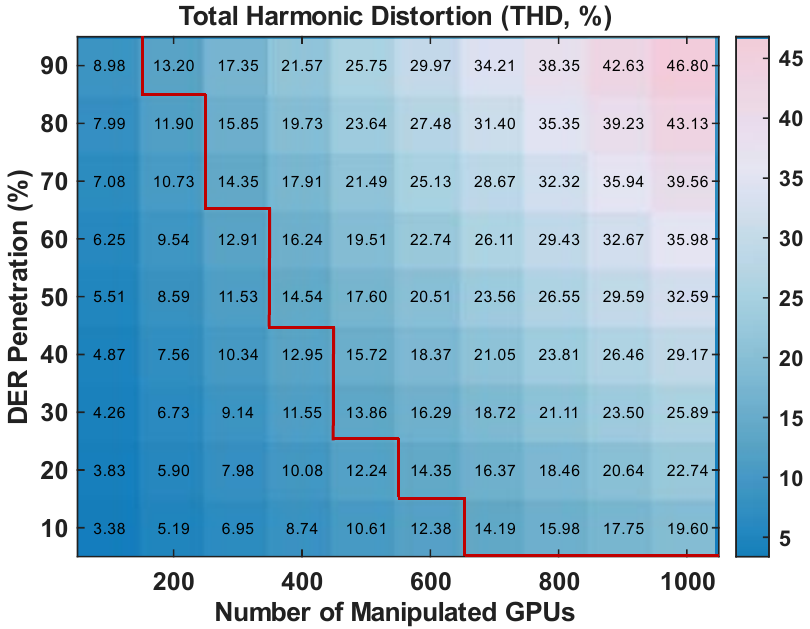}} 
\hspace{-0.9em} 
\subfigure[Instability (Damping ratio)]
{
        \label{damping_impact}
		\includegraphics[align=t,width=0.33\linewidth]{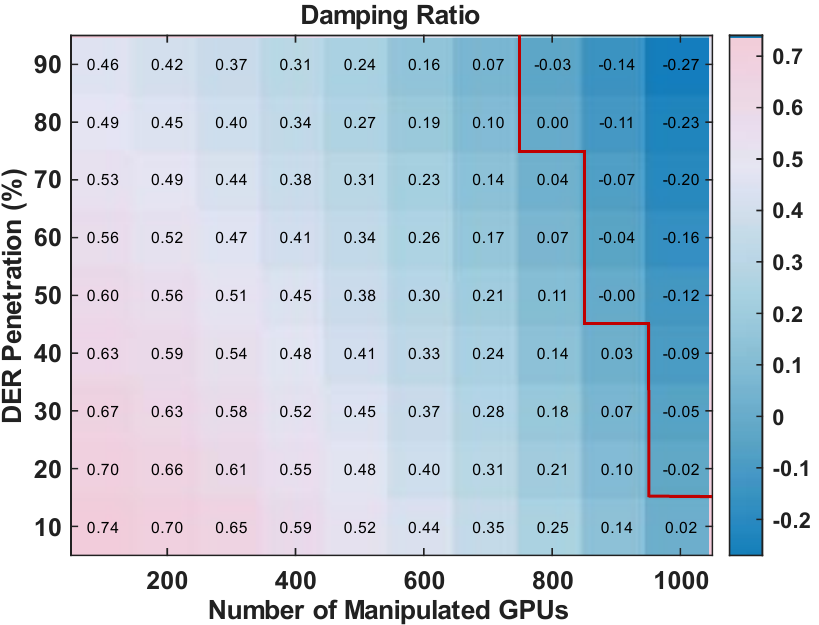}} \\
\subfigure[Current waveforms under different attacks]
{
        \label{simulation_current}
		\includegraphics[align=t,width=0.45\linewidth]{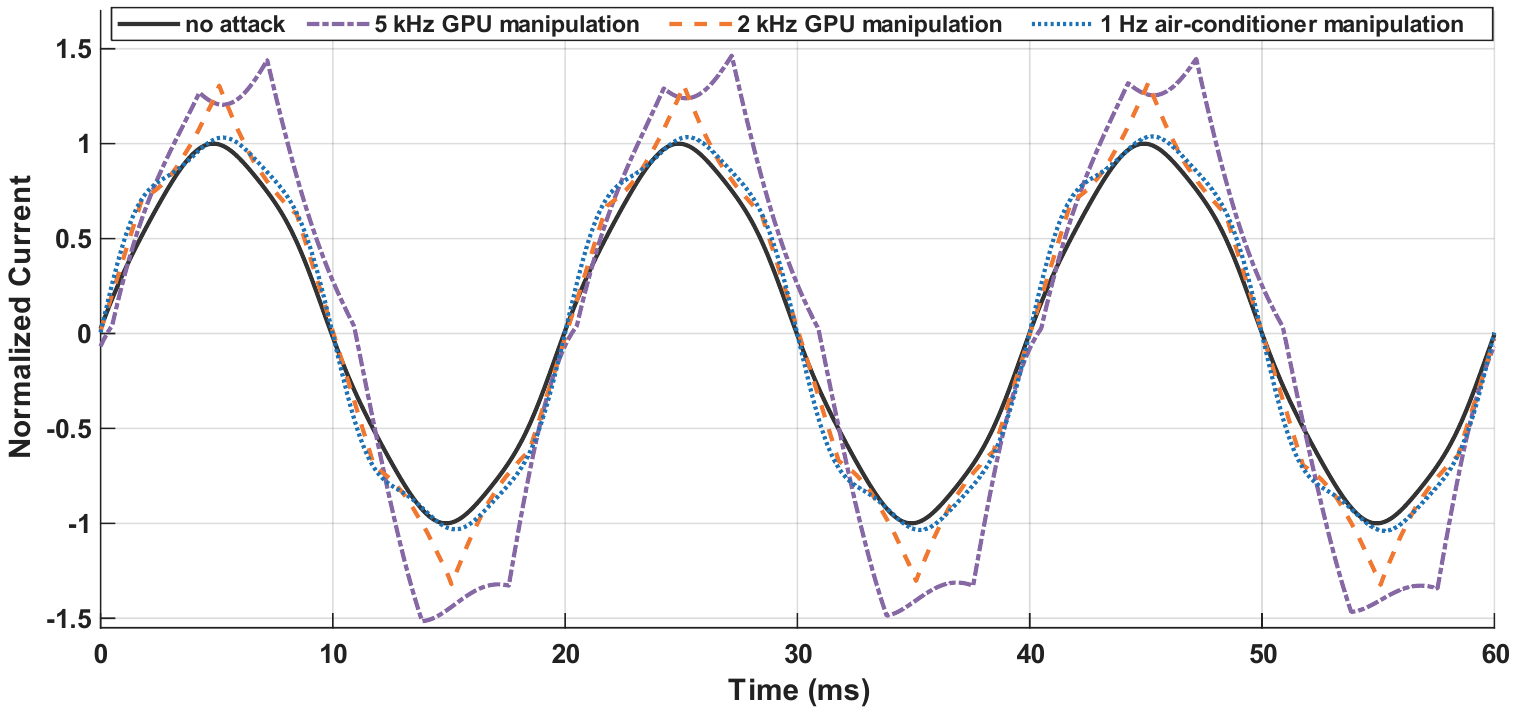}} 
\hspace{-0.9em} 
\subfigure[Harmonic spectrum under different attacks]
{
        \label{simulation_FFT}
		\includegraphics[align=t,width=0.455\linewidth]{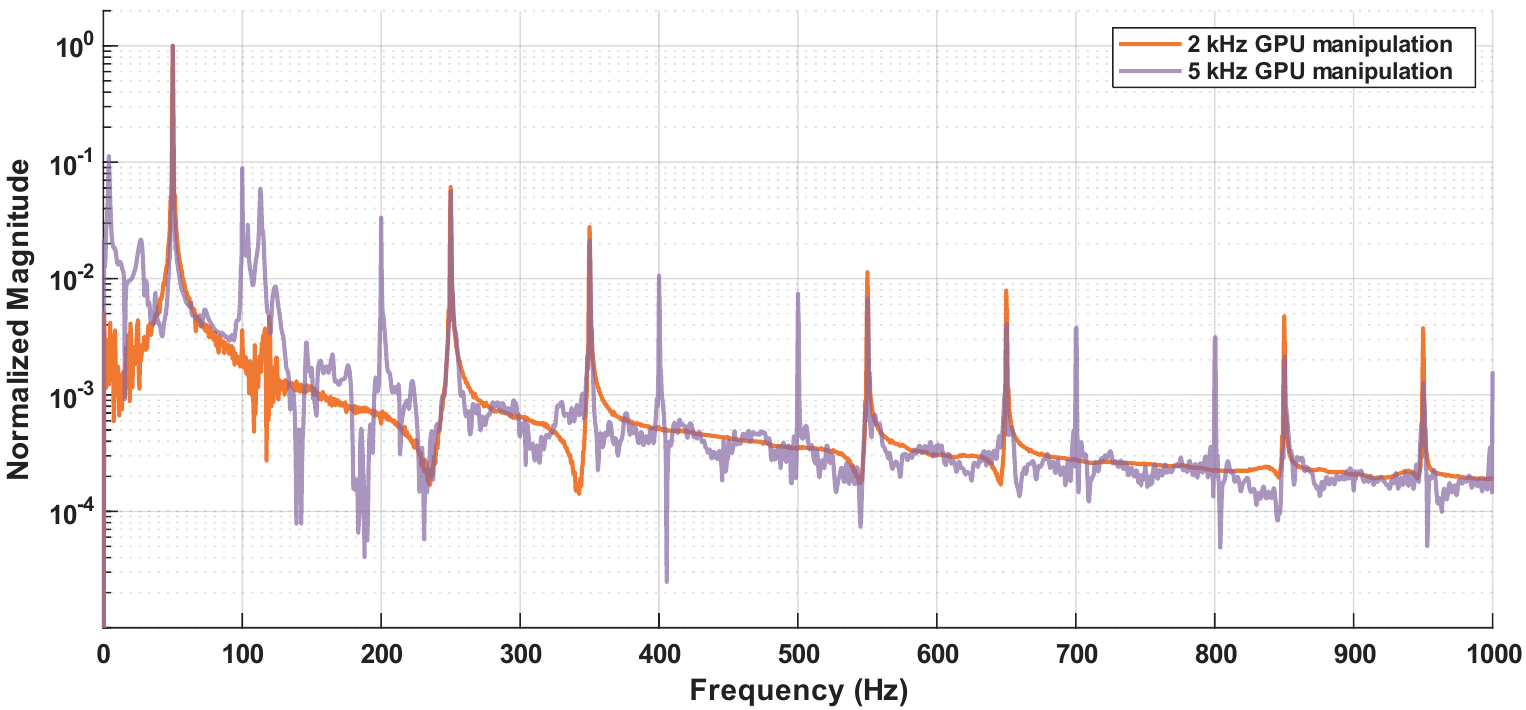}} \\
        \vspace{-0.1in}
\subfigure[Low-frequency oscillation induced by \texttt{Bit2Watt} risks. The attack is launched at 1 s and the grid is separated into two areas according to frequencies.]
{
        \label{fre_oscilation1}
		\includegraphics[align=t,width=0.8\linewidth]{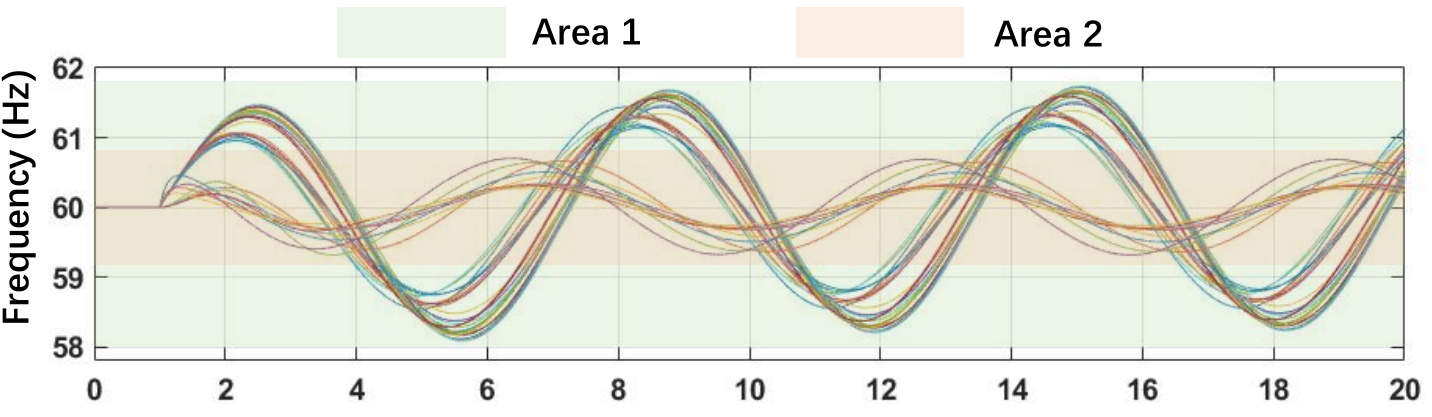}}
        \vspace{-0.1in}
\caption{Impact of \texttt{Bit2Watt} risks on a 1-MW local power system. The red lines in (a) (b) (c) mark the standard-specified thresholds, separating values above and below them.}
\label{impact}
\end{figure}

\section{Experiments}
We assess the impact of \texttt{Bit2Watt} on local power infrastructure and investigate the scale at which the GPU power manipulation leads to tangible physical effects. We begin with simulation-based analysis to characterize the system-level response across different attack configurations. Guided by the results, we conduct real-world experiments on a locally deployed power system to corroborate simulation findings and confirm the physical realizability.

\subsection{Simulation Evaluation}

\textbf{GPU load aggregation setup.} We adopt the NVIDIA RTX 3090 as the reference GPU model, and simulate a local power system with an aggregate load of 1 MW, of which approximately one quarter consists of the victim GPU loads and the remaining three quarters comprise background loads such as household appliances. This composition is motivated by electricity consumption patterns in Virginia, which hosts the world’s highest concentration of data centers \cite{IDCproportion}. The simulation information can be found in Figure \ref{SimulationTopo}, Table \ref{tab:impedance} and Table \ref{tab:control_settings}.

To evaluate the worst-case impact on the grid, the GPU units are aggregated under a synchronized assumption, where all units operate with identical modulation frequency and phase. In Simulink, this aggregated behavior is emulated by a controlled current source driven by a \texttt{MATLAB Function} block, defined as
\[
I_{\mathrm{GPU}}(t)=-\min\!\left(\frac{P_{\mathrm{GPU}}(t)}{\max(|V_{\mathrm{SPS}}|,80)},\ 10\right)\ \text{A},
\]
where
$P_{\mathrm{GPU}}(t)=50+200\sin(2\pi f_{\text{att}}\, t)\ \text{kW}$,
and $V_{\mathrm{SPS}}$ denotes the local DC voltage used for scaling the load current.
Device-level heterogeneity, phase dispersion, and communication delays across individual GPU units are not explicitly modeled and thus the reported results provide a conservative bound on realistic system behavior.

\textbf{Result analysis.} The simulation results in Figures \ref{V_impact}, \ref{THD_impact}, and \ref{damping_impact} collectively demonstrate that coordinated GPU power manipulation can simultaneously induce the voltage excursion, harmonics, and instability of a local power system, with these effects being significantly exacerbated by high levels of DER penetration. Taking a 50\% DER penetration level as an example, manipulating only 500, 400, and 900 GPUs is sufficient to degrade voltage quality, increase current harmonic distortion, and compromise system stability beyond the thresholds specified by relevant standards \cite{IEEE1547, IEC61000_3_12}.

Figures~\ref{simulation_current} and \ref{simulation_FFT} show the current waveforms and harmonic spectra under 50\% DER penetration with 1,000 GPUs manipulated. GPU power modulation induces severe current distortion, producing a sharp, peaky waveform with a high crest factor that increasingly deviates from the nominal sinusoid as the attack frequency rises. In contrast, low-frequency disturbances analogous to air-conditioner loads (blue dotted lines) produce negligible effects. Physically, the observed pulsating current reflects the GPU cluster’s rapid power ramps, which draw short, pulse-like currents during fast load transitions.

In Figure \ref{simulation_FFT}, a distribution of high-order harmonics is observed and a persistent cluster of high-frequency components extends across the spectrum, contributing to the substantial rise in total harmonic distortion (THD). Under a 90\% DER penetration level, the calculated THD reaches an extreme value of 46.80\% when 1000 GPUs are manipulated. Even at a lower penetration of 10\%, the THD remains as high as 19.60\%, both of which substantially exceed the 13\% guideline recommended by IEC 61000-3-12 \cite{IEC61000_3_12}.

More critically, at 90\% DER penetration, the damping ratio drops from 0.46 with 200 manipulated GPUs to –0.27 with 1,000 GPUs, marking a transition from a well-damped to an unstable regime. This instability emerges as the NIR induced by coordinated GPU power modulation progressively overwhelms the inherent damping of the grid and inverter controls. A critical threshold appears at approximately 800 manipulated GPUs under 80\% DER penetration, beyond which the damping ratio becomes negative and the system enters an oscillatory regime.

Based on the Simulink results, we further examine area oscillations using the IEEE 118-bus system (Figure~\ref{118bus}). Under attacks, the system exhibits oscillatory behavior driven by GPU-induced power variations. As the damping ratio crosses zero, the response transitions from stable to oscillatory.
As shown in Fig.~\ref{fre_oscilation1}, this leads to frequency bifurcation, where the network splits into two coherent frequency clusters. The resulting loss of synchronism indicates transient grid partitioning, which may trigger widespread protection actions and potentially lead to a black-start condition.

\subsection{Real-world Experiments}

\begin{figure} [t]
\centering
\subfigure[System topology]
{
        \label{LowFilterTopo}
		\includegraphics[align=t,width=0.45\linewidth]{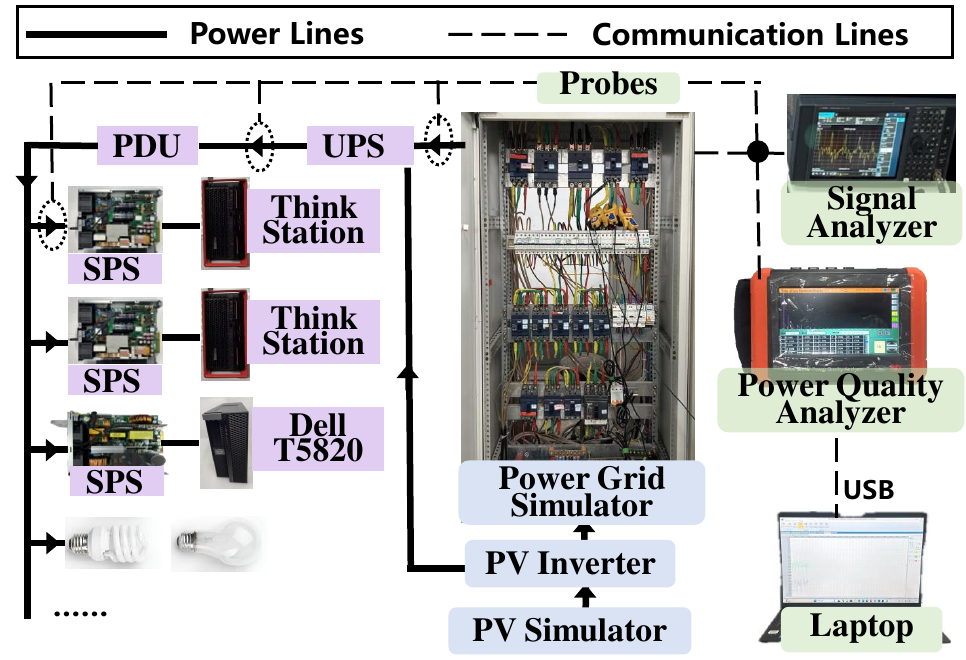}} 
\subfigure[Experiment testbed]
{
        \label{low-filter-setup}
		\includegraphics[align=t,width=0.46\linewidth]{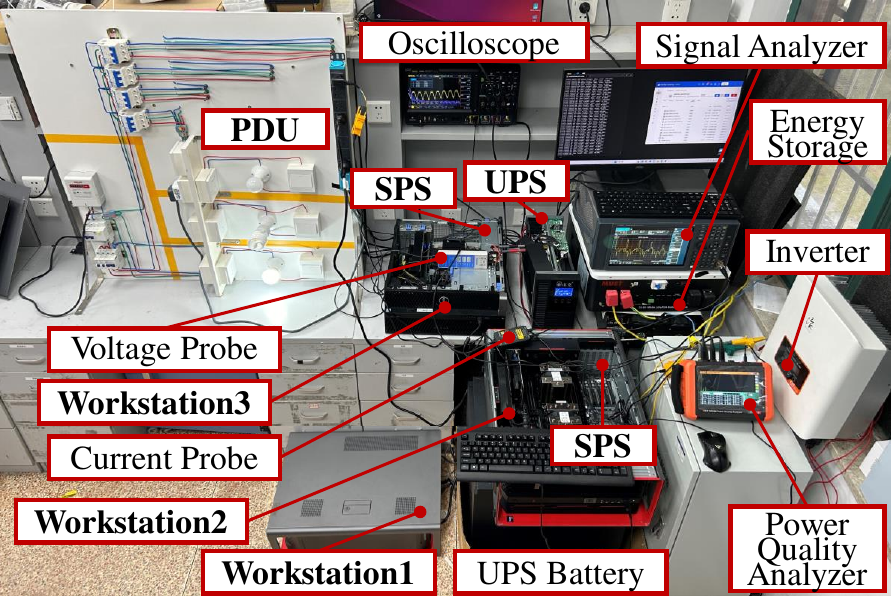}} \\
        \vspace{-0.05in}
\subfigure[SPS input current]
{
        \label{PSUcurrent}
		\includegraphics[align=t,width=0.32\linewidth]{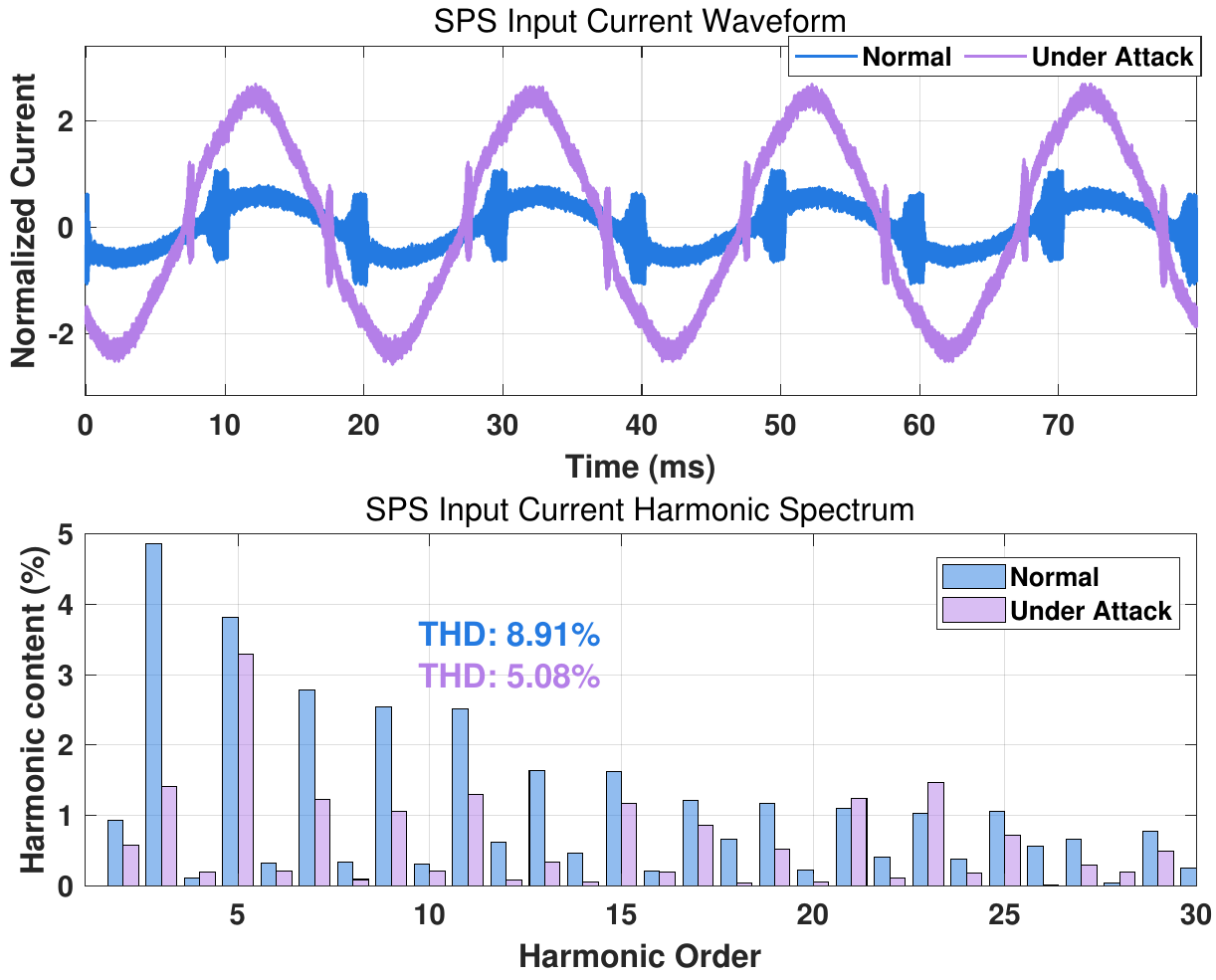}} 
\hspace{-0.9em} 
\subfigure[PDU input current]
{
        \label{PDUcurrent}
		\includegraphics[align=t,width=0.32\linewidth]{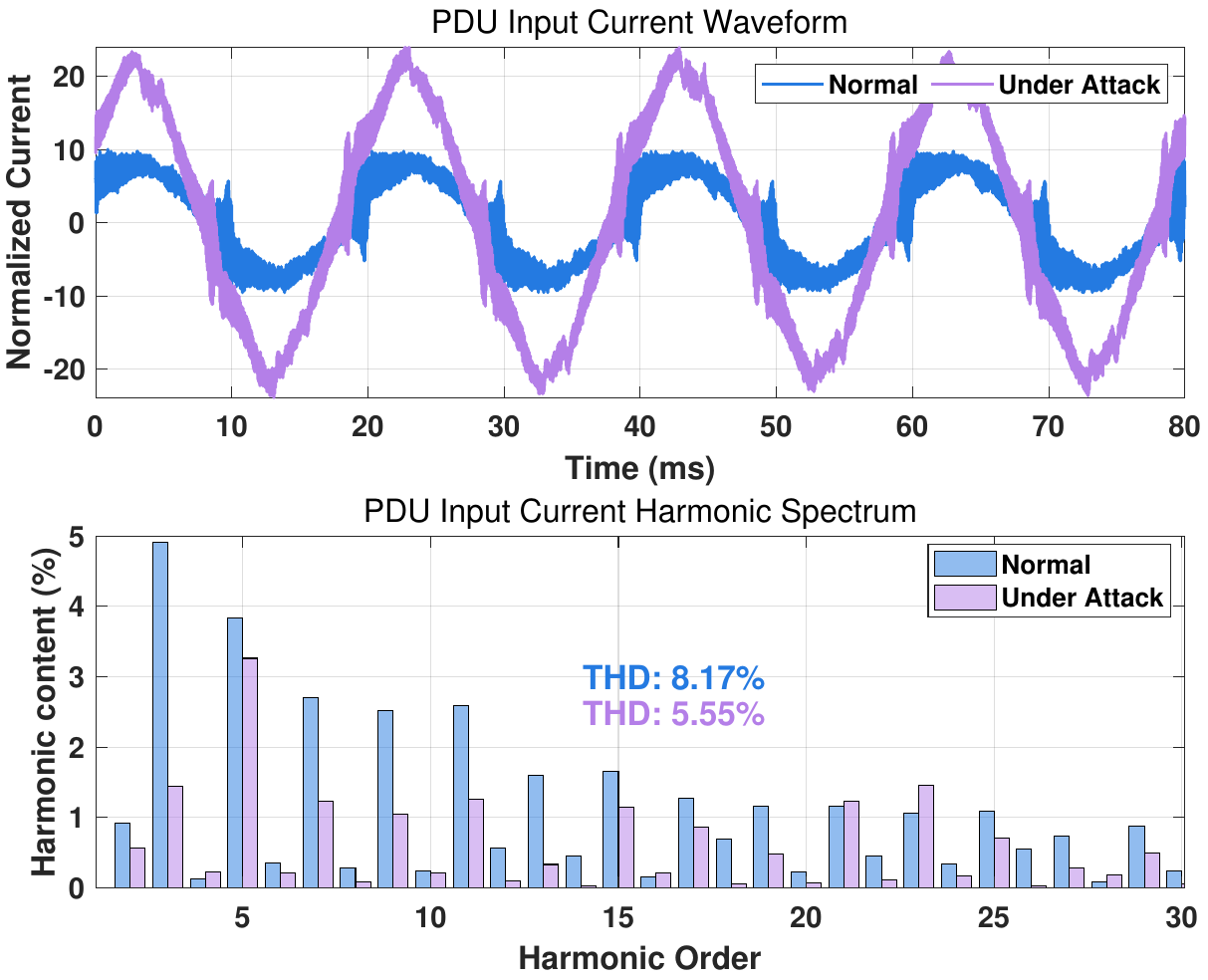}}
\hspace{-0.9em} 
\subfigure[UPS input current]
{
        \label{UPScurrent}
		\includegraphics[align=t,width=0.32\linewidth]{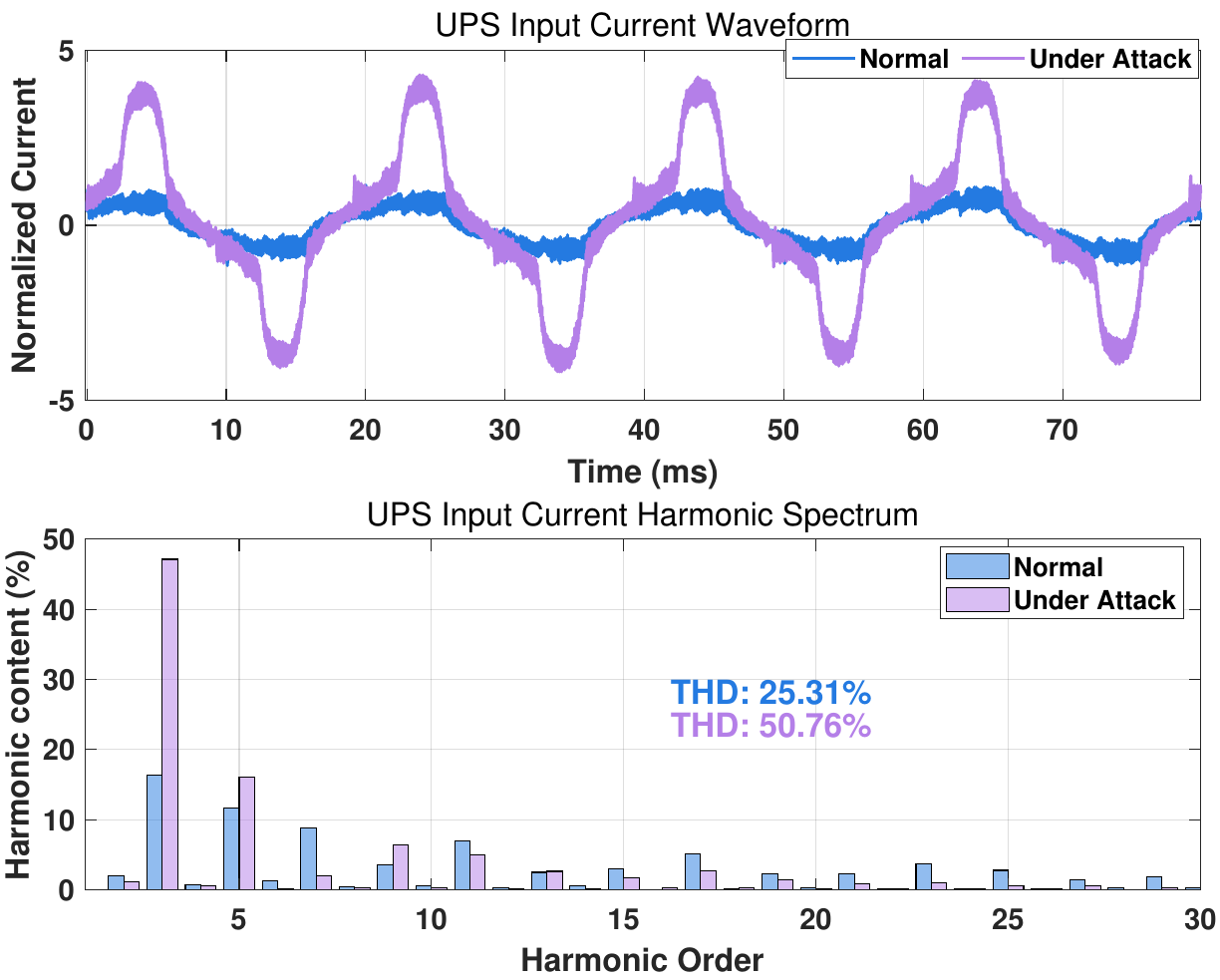}}
        \vspace{-0.1in}
\caption{Results of the hardware experiments.}
	\label{realworld_experiment}
\end{figure}

\begin{figure} [t]
\centering
\subfigure[Stage effect on attack-induced spectral change]
{
        \label{low-filter-fft}
		\includegraphics[align=t,width=0.45\linewidth]{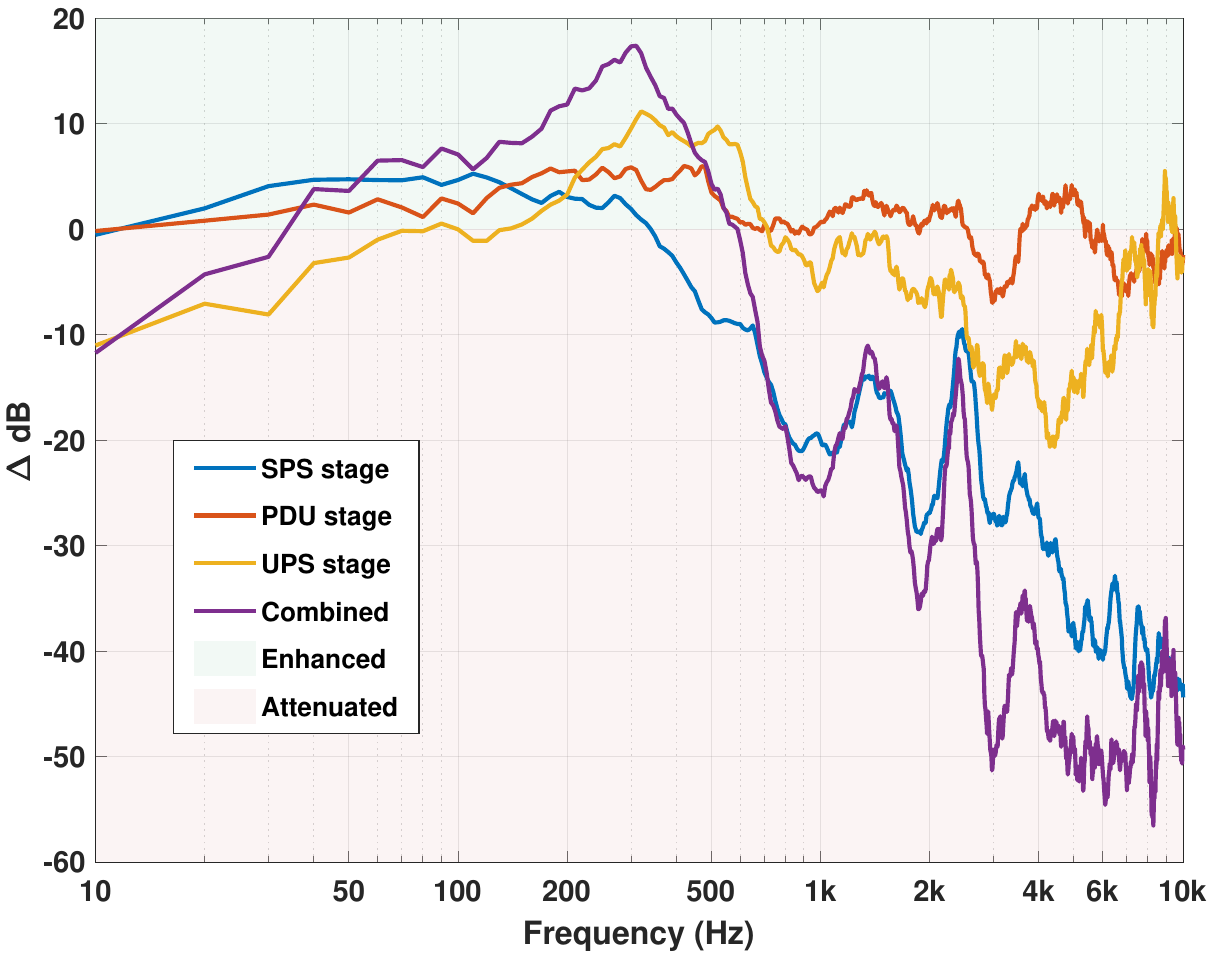}}
\subfigure[UPS DC bus voltage]
{
        \label{UPSvoltageDCbus}
		\includegraphics[align=t,width=0.455\linewidth]{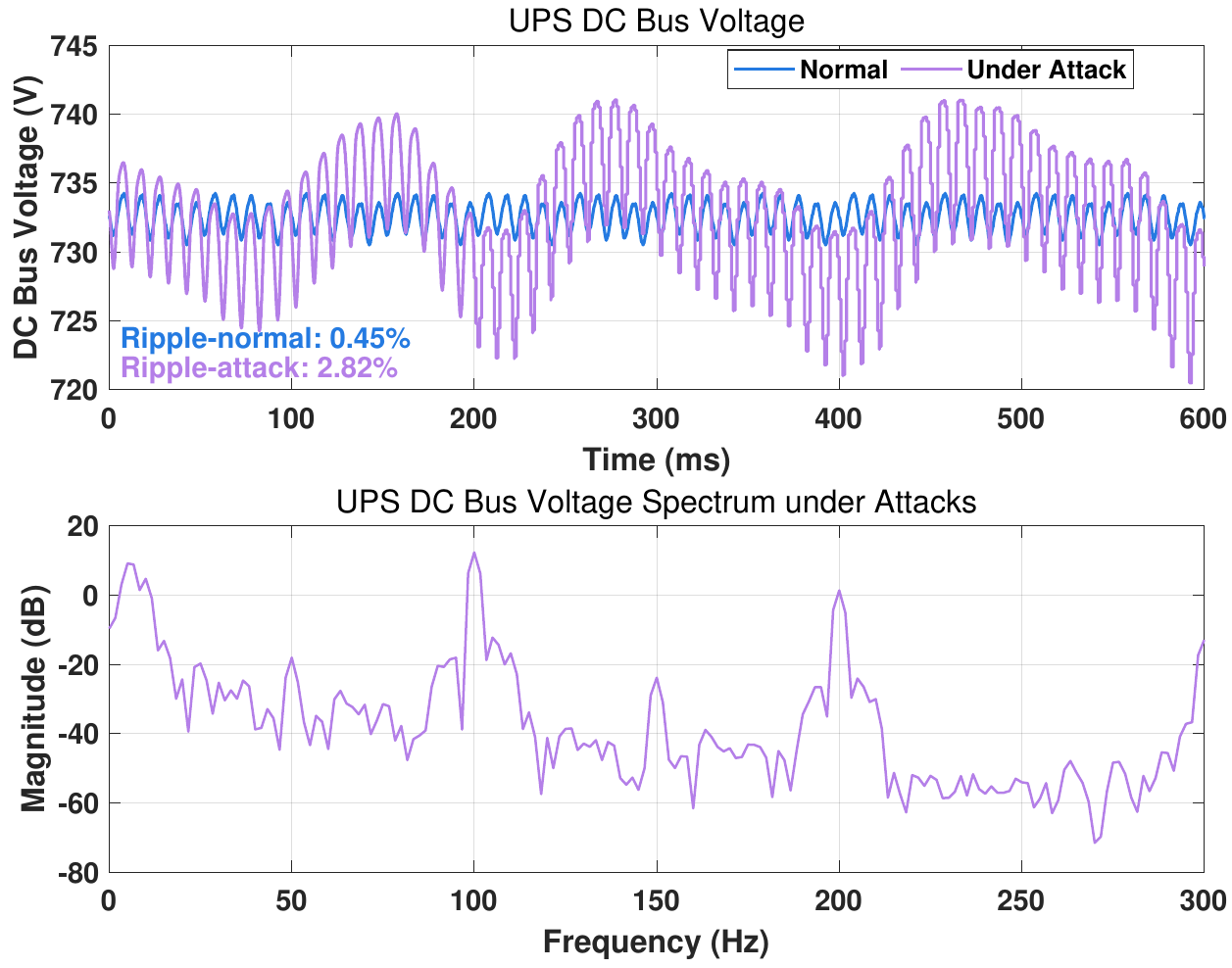}}
        \vspace{-0.1in}
\caption{Spectral attenuation and reshaping across the power delivery chain.}
	\label{low-filter_experiment}
\end{figure}

\textbf{Experiment setup.} 
We build a local power grid testbed that emulates a small-scale, inverter-dominated distribution system, as illustrated in Figure \ref{LowFilterTopo}. The system consists of a power grid simulator that provides a 220 V AC supply, a PV simulator, a grid-tied PV inverter, an energy storage battery, and a diverse set of electrical loads. These loads include typical household appliances (e.g., water heaters, bulbs) as well as three workstations with six GPUs. The total power demand of the testbed is approximately 8 kW, with roughly 40\% supplied by the PV units, resulting in a mixed grid–DER operating condition. Electrical measurements are collected and analyzed using a handheld power quality analyzer (VICTOR 5000M) and KEYSIGHT EXA Signal Analyzer N9010B.

The prototype power supply chain of the workstations consists of GPU–SPS–PDU–UPS– PCC, as shown in Figure \ref{low-filter-setup}. The setup includes two ThinkStation PX units powered by dedicated SPS modules (Delta DPS-1850CB A, 1850 W), and one Dell T5820 powered by an SPS module (Dell H950EF-00, 950 W). These loads are aggregated at the PDU, where additional equipment is connected to emulate background disturbance loads. The PDU is implemented using an energy monitoring device (DELIXI DDS606), circuit breakers (DELIXI HDBE-63), and power strips. An AC UPS (SHANKE C3KS), equipped with eight 12 V battery cells connected in series, interfaces the system with the PCC.

\textbf{Ethical Considerations.}
All experiments were conducted in controlled environments. No attacks were performed on production systems, and no vulnerabilities in specific commercial products are disclosed.

\textbf{Result analysis.}
Figures \ref{PSUcurrent}, \ref{PDUcurrent}, and \ref{UPScurrent} compare current waveforms and harmonic spectra across the cascaded power delivery chain under normal and attack conditions. 
At all stages, the attack significantly increases the current magnitude. 
The THD at the SPS and PDU stages decreases, mainly due to the increased fundamental current and the low-pass filtering effect of the SPS. 
However, low-order harmonics (e.g., $3^{\text{rd}}$ and $5^{\text{th}}$) sharply increase at the UPS stage. 
This indicates that intermediate stages not only attenuate high-frequency components but also reshape the disturbance through control dynamics, ultimately projecting the distortion onto the upstream grid.

\subsubsection{Signal Propagation through Intermediate Power Electronics}

In this section, we quantify how the attack-induced spectral increment propagates across intermediate stages. For each stage $s$ and frequency $f$, we define
$
\Delta_s(f) := M_s^{\mathrm{atk}}(f) - M_s^{\mathrm{base}}(f),
$
where $M_s^{\mathrm{atk}}(f)$ and $M_s^{\mathrm{base}}(f)$ denote the spectral magnitudes (in dB) at stage $s$ under attack and attack-free conditions, respectively. Figure~\ref{low-filter-fft} shows the stage-induced variation of this increment, i.e.,
$
\Gamma_s(f) := \Delta_s(f) - \Delta_{s-1}(f),
$
where $s-1$ denotes the preceding stage. Hence, $\Gamma_s(f)>0$ indicates that stage $s$ enhances the attack-induced spectral increment at frequency $f$, whereas $\Gamma_s(f)<0$ indicates attenuation.

Figure~\ref{low-filter-fft} reveals a clear frequency-dependent behavior. The SPS stage introduces the strongest high-frequency attenuation, indicating that a large portion of the high-frequency modulation is absorbed locally. This is consistent with the SPS internal structure, where DC-bus capacitors and local power-conditioning circuits provide strong buffering and smoothing effects against fast fluctuations. In contrast, low-frequency components below approximately 300~Hz are mildly enhanced. The PDU stage has a relatively weak impact, consistent with its largely passive role.

The UPS stage exhibits mixed behavior, with attenuation at high frequencies but enhancement in the low-frequency range (200--700~Hz), indicating a reshaping effect. The complex UPS control system plays a decisive role here. Unlike the unidirectional regulation and straightforward control logic of an SPS, the UPS utilizes tightly coupled, cascaded feedback loops to synchronize the input rectifier with the output inverter. As shown in Figure \ref{UPSvoltageDCbus}, the attack induces pronounced low-frequency oscillations on the UPS DC bus voltage, increasing the ripple ratio from 0.45\% to 2.82\%, which exceeds the suggested limit \cite{VoltageRipple}. This is because rapid GPU power transients disrupt the nominal constant-power mode of the UPS and force it into current-limiting mode. It indicates that, although high-frequency components are suppressed upstream, the energy is converted into UPS DC-bus-level power fluctuations, which in turn distort the input current waveform.

Overall, the results support an attenuation-and-reshaping mechanism: the attack-induced spectral increment is selectively suppressed at high frequencies and reshaped at some lower-frequency components along the intermediate power electronics.

\subsection{Discussions}

\subsubsection{From Local to Wide-Area Power Infrastructure}

\begin{figure} [t]
	\centering
\subfigure[Cascading failure evolution]
{
        \label{Round13_9241_2}
		\includegraphics[align=t,width=0.42\linewidth, trim = 0 0 0 16, clip]{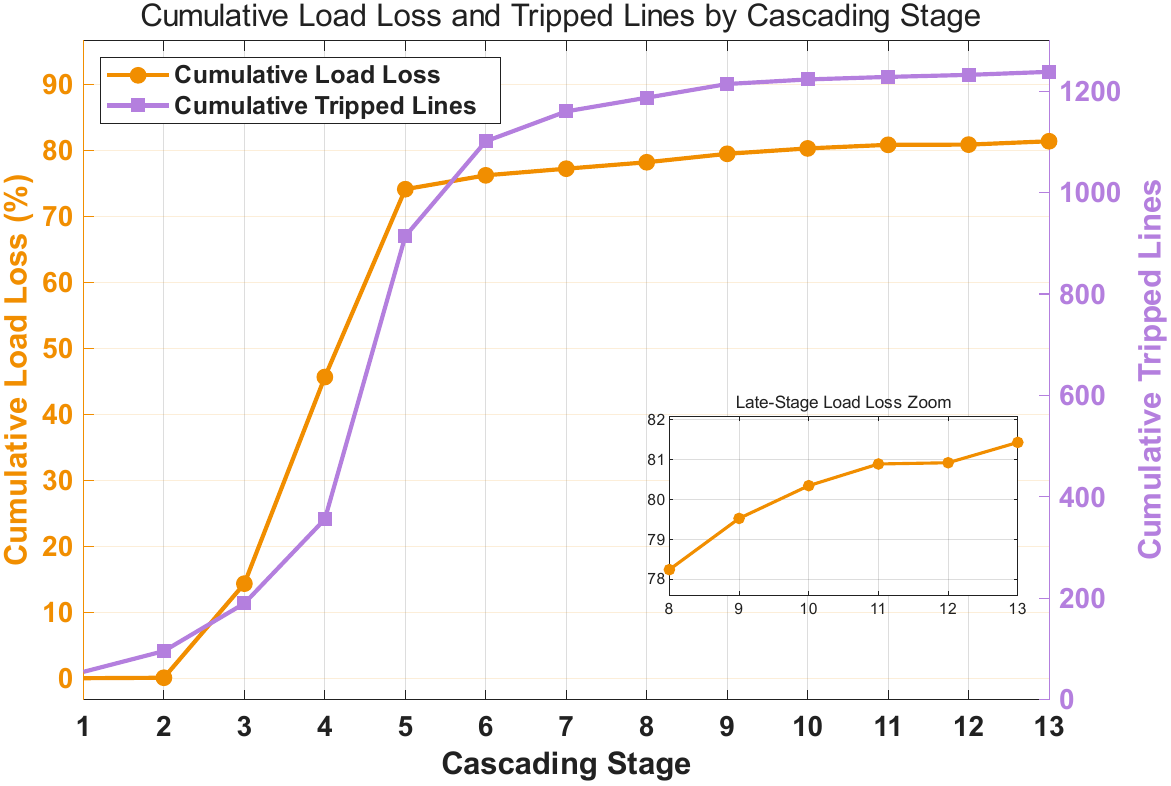}}
\hspace{2mm}
\subfigure[Impact of the chaotic effects]
{
        \label{Abruptness}
		\includegraphics[align=t,width=0.43\linewidth]{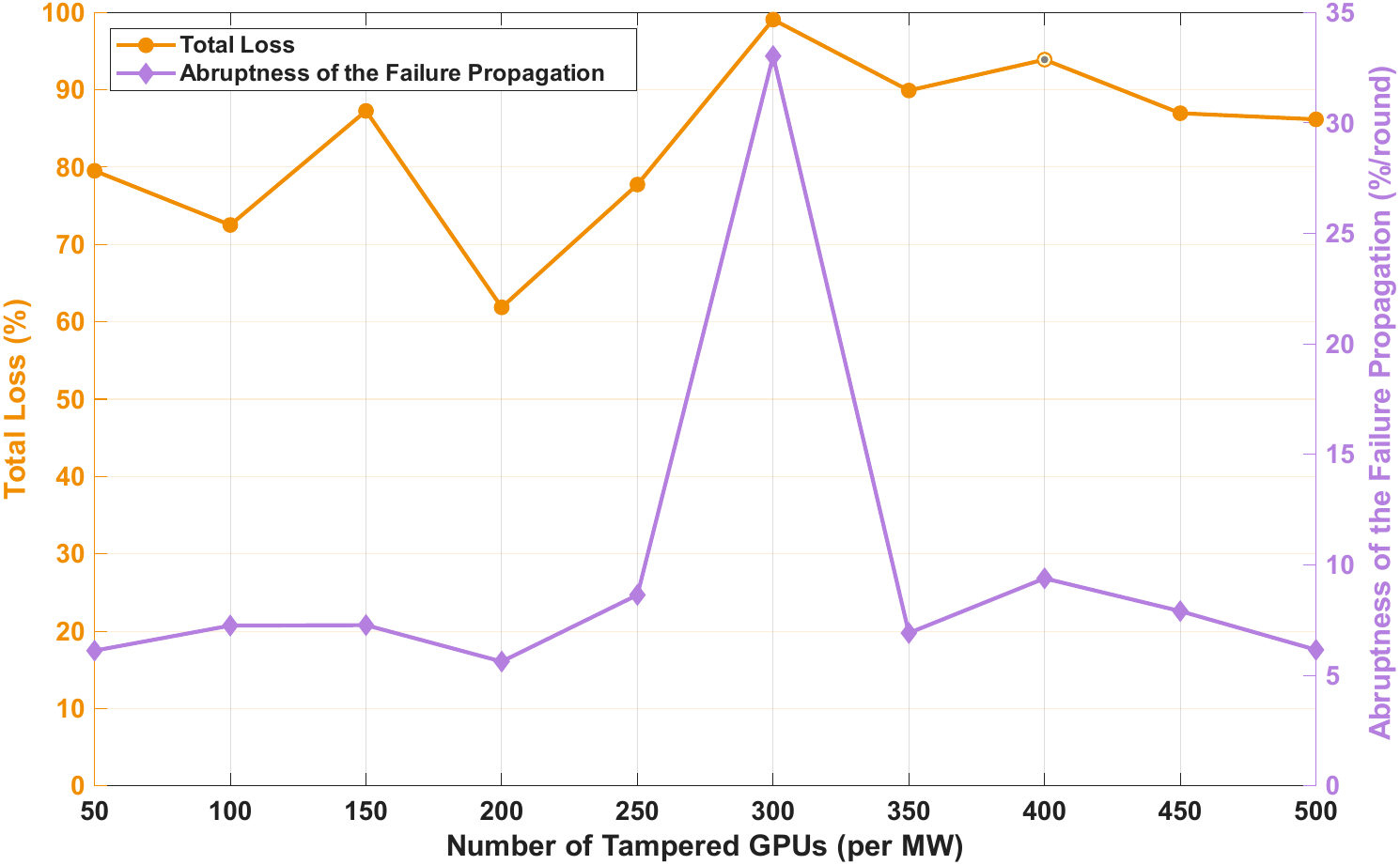}}
	\caption{Cascading results in the 9241-bus European transmission network and associated chaotic effects. }
	\label{Cascading}
\end{figure}

We extend the analysis from individual data-center interfaces to a large-scale transmission network with simulations on a modified 9241-bus European transmission network representative of modern DER-rich grids \cite{9241}. As shown in Figure~\ref{Round13_9241_2}, a localized disturbance (accounting for only 2\% of the total system load) triggers a cascade over 13 stages. Both the number of tripped lines and load loss increase sharply in the early stages, indicating a fast propagation phase, followed by a slower saturation regime. In total, 1,238 transmission lines are disconnected, leading to severe network fragmentation and a final load loss of approximately 81\%. 
These results demonstrate that disturbances originating at localized interfaces, when coupled with protection actions, can propagate through interconnected networks and result in large-scale cascading failures.

\subsubsection{Influence of Noisy Environments and Chaotic Effects}
\label{sec:noise}

The \texttt{Bit2Watt} attack may be degraded by environmental noise, which affects the synchronization of distributed GPUs, as timing jitter directly determines the efficacy of constructive power-wave superposition. We model the asynchronization $\Delta \tau_i$ of each victim GPU as a Gaussian random variable, i.e., $\Delta \tau_i \sim \mathcal{N}(0, \sigma_\tau^2)$, capturing the combined effects of network latency and OS scheduling jitter.
Such perturbations reduce the aggregated attack amplitude due to imperfect phase alignment. For a 2~kHz modulation, when $\sigma_\tau < 50\,\mu\text{s}$, the reduction in attack amplitude is limited, although the waveform gradually transitions from a square-like shape to a more sinusoidal profile. When $\sigma_\tau = 100\,\mu\text{s}$, the attack amplitude decreases by approximately 20\%. 
The corresponding waveforms are shown in the appendix.
The robustness of the attack decreases as the frequency increases.
However, this loss in aggregation efficiency caused by asynchronization can be compensated by increasing the number of victim devices, owing to the scaling effect of large populations.

Beyond external noise, the power-electronic-dominated system exhibits strong nonlinear dynamics that may lead to bifurcations and chaotic effects. In such regimes, the system can appear non-deterministic in practice and become sensitive under certain settings. To quantify the abruptness of failure propagation, we introduce a cascading progression rate, defined as the ratio between total load loss and the number of cascading rounds, which correlates with the temporal evolution of the cascade. Lower abruptness implies a slower progression, providing operators with more time for mitigation.
As shown in Figure \ref{Abruptness}, both the total load loss and the propagation abruptness exhibit sharp extrema when the number of tampered GPUs reaches 300/MW, indicating strong sensitivity to the initial attack configuration. The cascading process also shows irregular and unpredictable line disconnections, reflecting chaos-like behavior near the stability boundary.
A similar sensitivity is observed with respect to attack timing, which alters the initial system states (e.g., voltage magnitudes and phase angles). Under such conditions, the system may exhibit pseudo-stable behavior, such as bounded oscillations (Fig.~\ref{fre_oscilation1}) that appear stable but do not converge to an equilibrium.
This variability highlights a double-edged effect of the underlying nonlinear dynamics: near transition boundaries, small perturbations can trigger disproportionately large instability, while the same complexity limits the attacker’s ability to optimize attack strategies based on partial system knowledge.

\color{black}
\subsubsection{Influence of System Parameters: Inertia and Damping}

\begin{figure} [t]
	\centering
    \hspace{5mm}
    \subfigure[Peak frequency]{
		\label{peakfrequency}
		\includegraphics[align=t,width=0.35\linewidth]{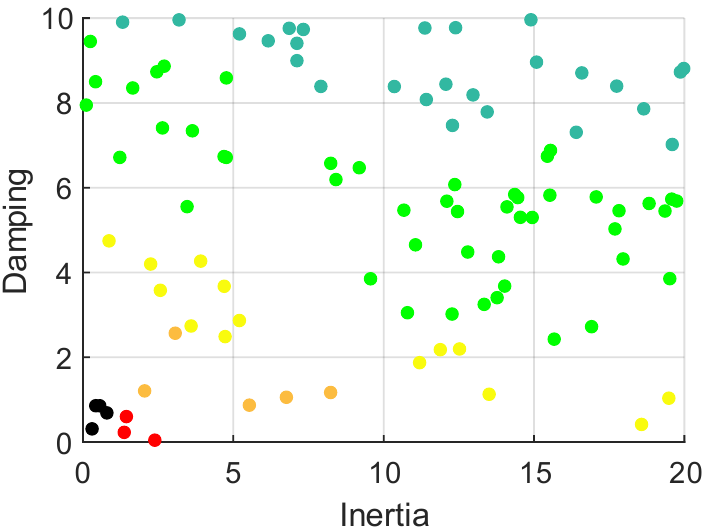}}
	\hfill
    \hspace{-5mm}
    \subfigure{
        \label{colorbar1}
        \includegraphics[align=t,width=0.054\linewidth]{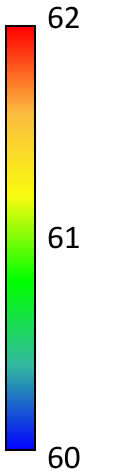}}
    \hspace{13mm}
    \setcounter{subfigure}{1}
    \subfigure[Steady frequency]{
        \label{stablefrequency}
        \includegraphics[align=t,width=0.35\linewidth]{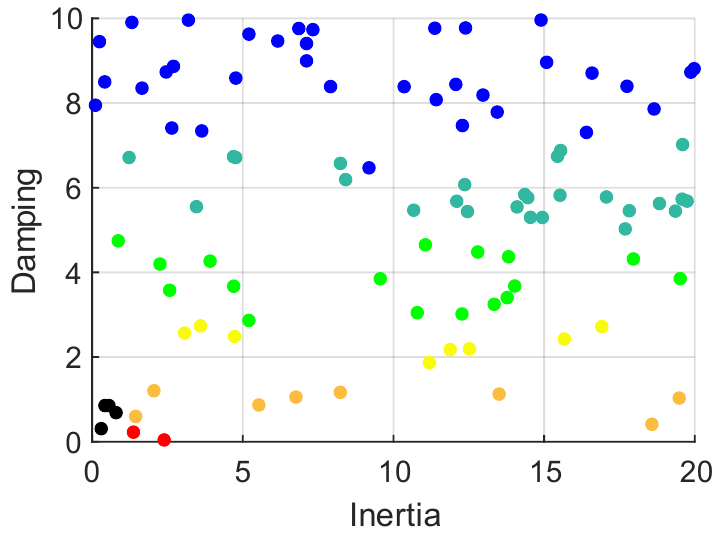}}
    \hspace{-3mm}
    \subfigure{
        \label{colorbar2}
        \includegraphics[align=t,width=0.054\linewidth]{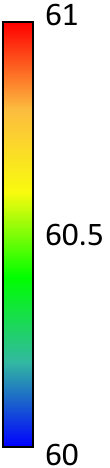}}
	\caption{\textcolor{black}{Frequency as a function of inertia and damping. The frequency values are color-coded, with cool colors representing low values and hot colors indicating high values. Black indicates that the grid becomes destabilized under this combination.}}
	\label{inertia_damping}
\end{figure}

Inertia fundamentally governs the system’s immediate frequency response by determining how rapidly frequency deviates following a disturbance, a role that becomes increasingly critical as high DER penetration displaces synchronous generation. Damping dictates the system’s ability to attenuate post-disturbance oscillations and restore stable operation.

To examine their relative influence, we conduct simulations on the IEEE 118-bus system using 100 different inertia–damping combinations. Figure\textcolor[RGB]{50,205,50}{~\ref{inertia_damping}} shows the peak and steady-state frequency following the attack. Increasing either inertia or damping improves frequency stability. 
Notably, damping exhibits a stronger stabilizing effect than inertia. As illustrated in Figure\textcolor[RGB]{50,205,50}{~\ref{peakfrequency}}, the peak frequency deviation decreases more rapidly along the damping axis than along the inertia axis. This indicates that, under GPU power manipulations, system stability is more sensitive to damping degradation than to inertia reduction, highlighting the critical role of damping in inverter-dominated power systems.

\subsubsection{Detectability and Stealth Analysis}
\begin{figure} [t]
\centering
\subfigure[Power observability across different power monitoring tools during \texttt{Bit2Watt}.]
{
        \label{PowerMonitorData}
		\includegraphics[align=t,width=0.46\linewidth]{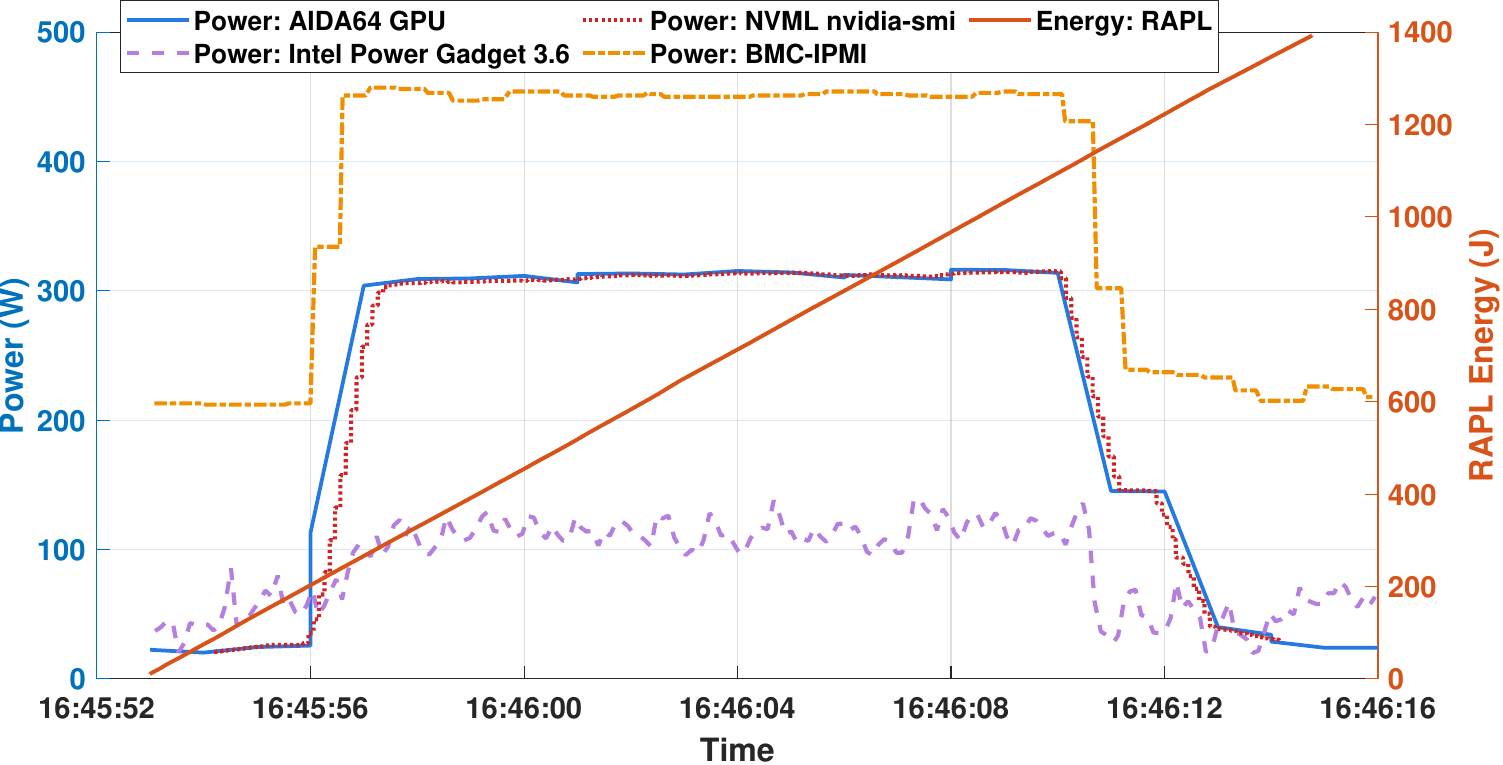}}
\hspace{2mm}
\subfigure[Detectability of \texttt{Bit2Watt} based on four kinds of monitoring settings]
{
        \label{detection_method}
		\includegraphics[align=t,width=0.425\linewidth]{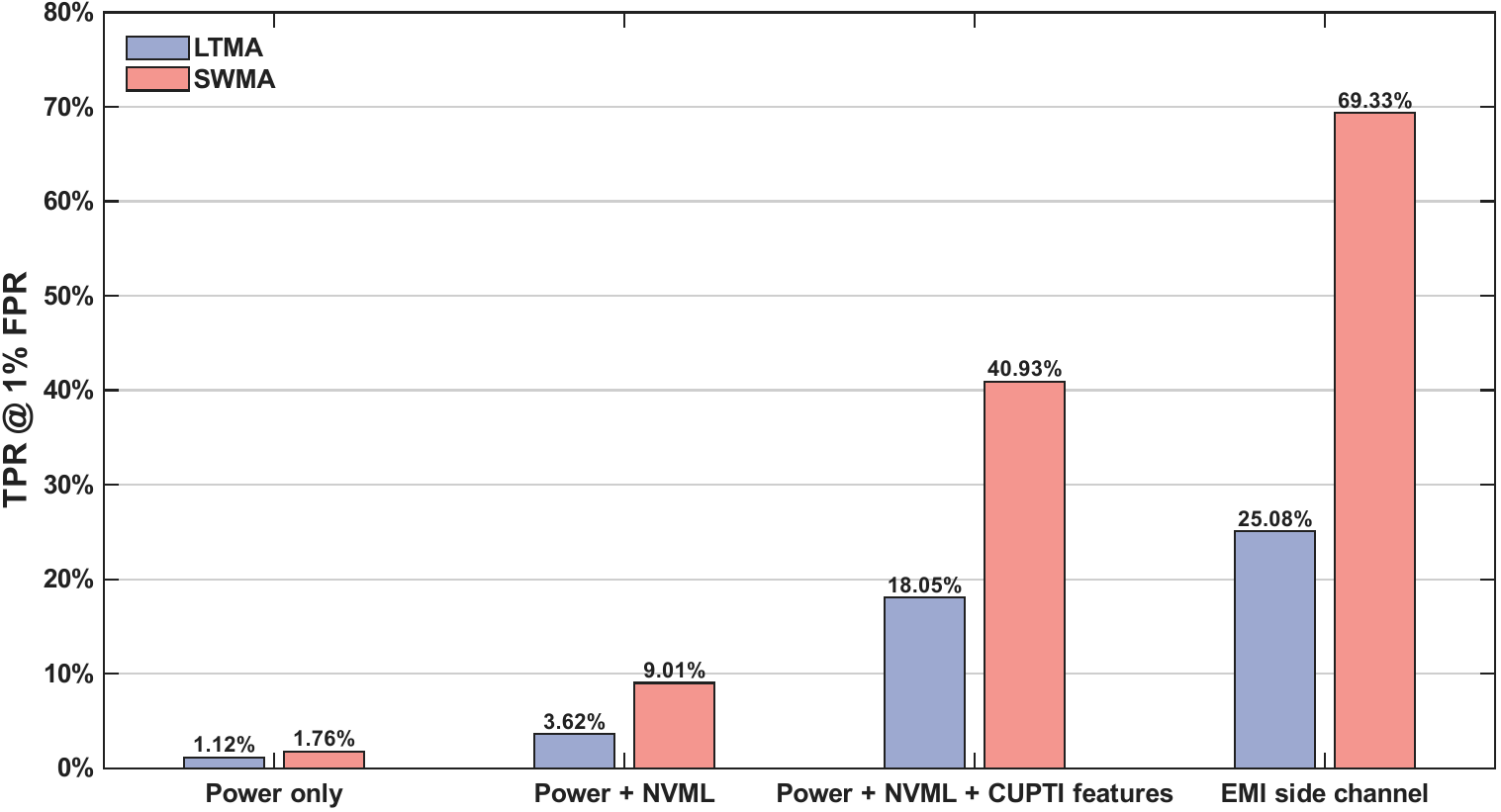}}
\caption{Detectability and stealth evaluation.}
	\label{detection}
\end{figure}

To evaluate the detectability and stealth under realistic monitoring, we conduct the quantitative study summarized in Table~\ref{tab:power_telemetry_interfaces} and Figure~\ref{detection}. Table~\ref{tab:power_telemetry_interfaces} reviews representative cloud- and facility-side monitoring interfaces, including NVML, RAPL, Intel Power Gadget, BMC/IPMI, and rack-level PDU telemetry. These interfaces are primarily designed for operational monitoring and logging rather than adversarial detection, and they typically expose coarse-grained, temporally averaged statistics instead of high-bandwidth instantaneous waveforms. Figure~\ref{PowerMonitorData} corroborates this observation empirically: the \texttt{Bit2Watt} signature is progressively smoothed from GPU-side telemetry to CPU-, server-, and rack-level monitoring. Since no established detector specifically targets this attack class, we implement a lightweight logistic-regression baseline and evaluate four settings: power-only monitoring, power combined with NVML runtime features, power combined with NVML and CUPTI features, and EMI side channel sensing from the power line and VRM. As shown in Figure~\ref{detection_method}, coarse cloud/facility-side telemetry provides only limited detectability, and LTMA is consistently harder to detect than SWMA under the same detector family. In contrast, dedicated physical-side sensing substantially improves detection. These results suggest that the mere availability of monitoring data does not directly translate into reliable detection of Bit2Watt-style attacks; rather, detectability depends critically on the monitoring layer, temporal granularity, and whether the defender deploys profiling-aware or physical-layer instrumentation beyond routine operational telemetry.

\begin{table}[t]
\caption{Typical power monitoring tools related to \texttt{Bit2Watt}.}
\label{tab:power_telemetry_interfaces}
\centering
\footnotesize
\setlength{\tabcolsep}{3pt}
\renewcommand{\arraystretch}{1.10}
\begin{tabularx}{\columnwidth}{
@{}>{\centering\arraybackslash}m{1.8cm} 
@{}>{\centering\arraybackslash}m{1.8cm}
@{}>{\centering\arraybackslash}m{2.4cm}
@{}>{\centering\arraybackslash}m{4.2cm}
C@{}}
\toprule
\makecell{\textbf{Tool}} 
& \makecell{ \textbf{Scope}} 
& \makecell{\textbf{Achieved Sampling} \\ \textbf{Frequency}} 
& \textbf{Related Data} 
& \makecell{\textbf{Access} \\ \textbf{Method}} \\
\midrule
AIDA64
& GPU, CPU
& 10 Hz
& Power, voltage, current
& User-level software \\

NVML
& GPU
& $450$ Hz
& Power, utilization, temperature
& \texttt{nvidia-smi} command \\

RAPL
& CPU
& 1 kHz
& Cumulative energy
& \texttt{powercap/intel-rapl}\\

Intel Power Gadget 3.6
& CPU
& 10 Hz
& Package power, frequency, temperature
& User-level GUI \\

BMC  & Server level  & 1 kHz  & Power, voltage, current
& \texttt{ipmitool} command \\

PDU  &  Rack level   & 1 Hz  & Power, energy, voltage, current  & User-level GUI\\
\bottomrule
\end{tabularx}
\end{table}

\subsubsection{\texttt{Watt2Bit} Risk}

\begin{figure} [t]
\centering
\subfigure[Additional thermal stress]
{
        \label{thermal}
		\includegraphics[align=t,width=0.33\linewidth, clip,trim=0 0 0 16]{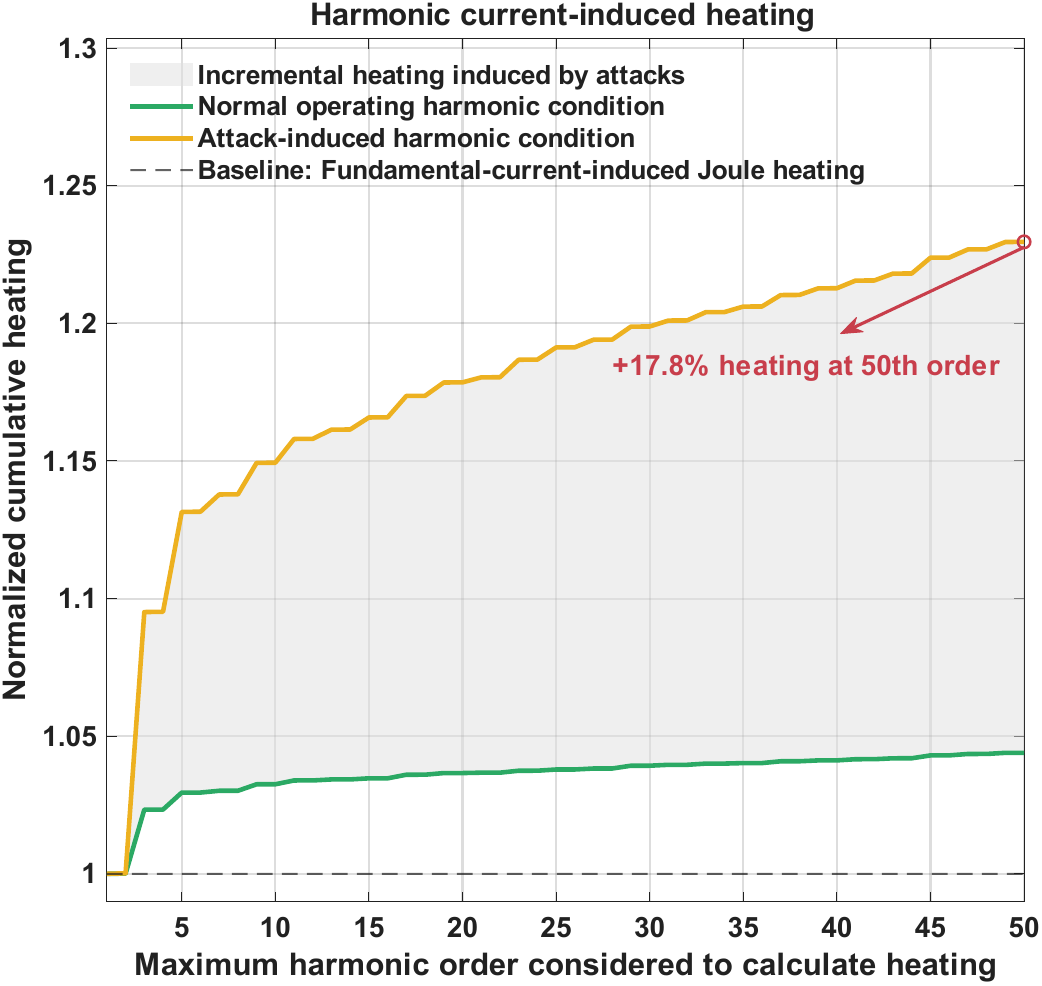}} 
\hspace{-0.9em} 
\subfigure[DoS due to overcurrent]
{
        \label{IEC60255}
		\includegraphics[align=t,width=0.33\linewidth, clip,trim=0 0 0 16]{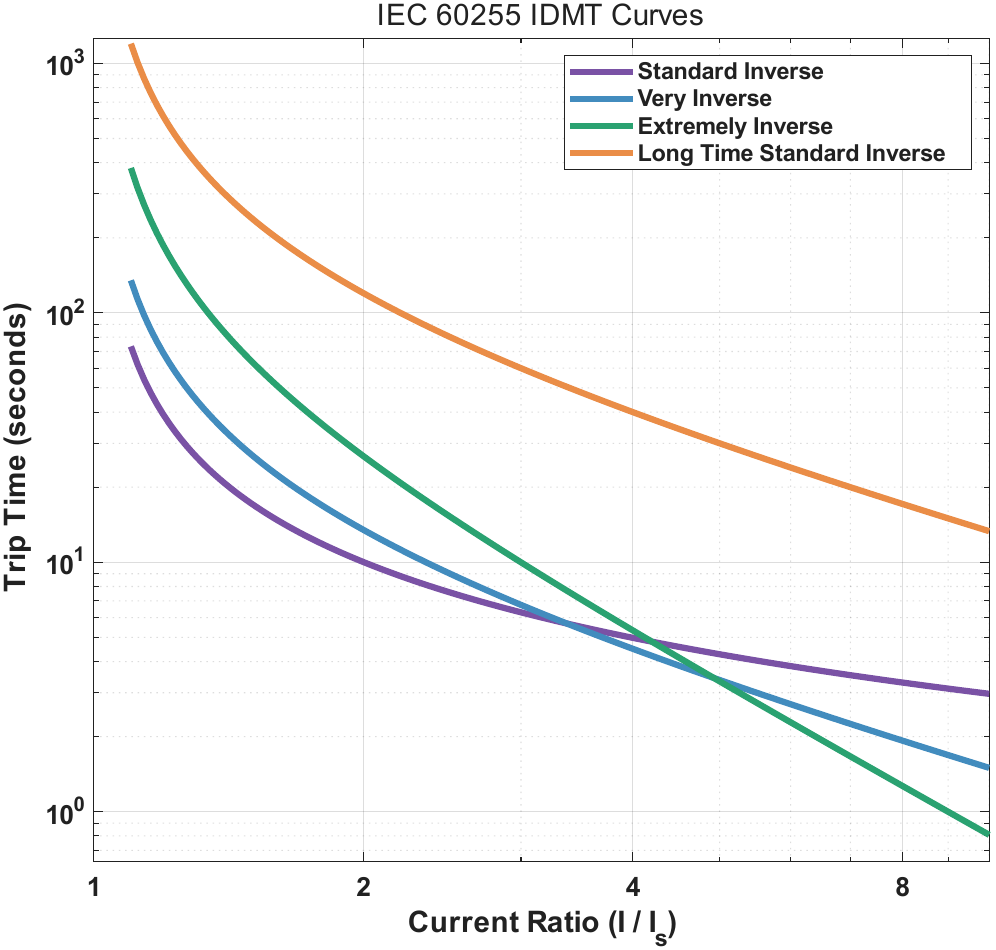}} 
\hspace{-0.9em} 
\subfigure[Information exfiltration]
{
        \label{exfiltration}
		\includegraphics[align=t,width=0.33\linewidth, clip,trim=0 0 0 16]{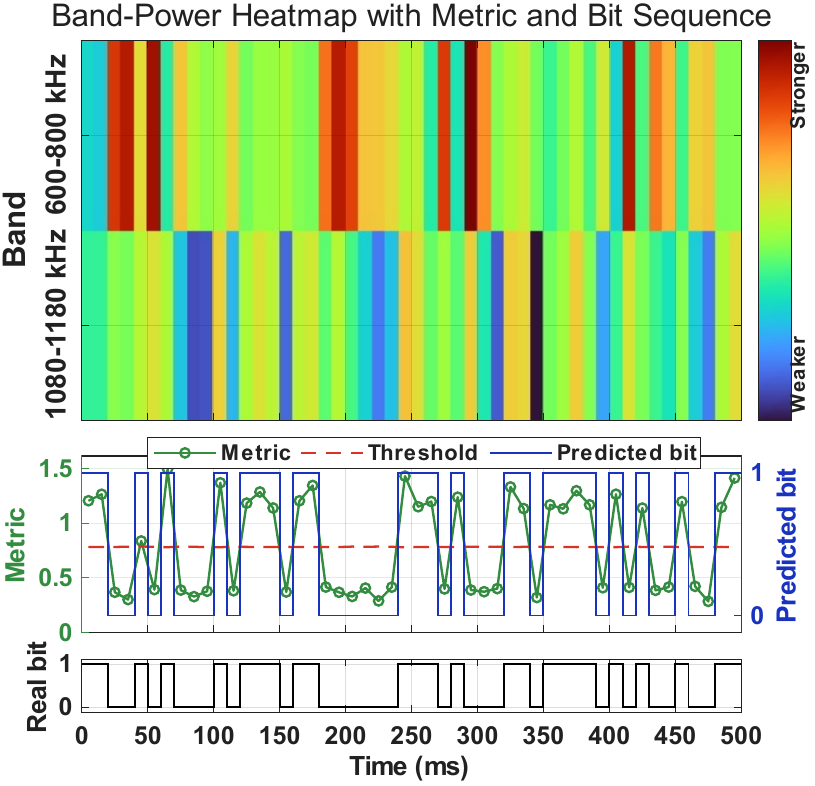}}
\caption{\texttt{Watt2Bit} risk analysis.}
\label{Watt2Bit}
\end{figure}

Figure~\ref{thermal} presents the normalized cumulative heating under nominal and attack conditions. Under normal operation, the cumulative heating remains close to unity, indicating that the contribution of higher-order harmonics is limited. In contrast, the attack-induced spectrum produces a clear and progressively widening deviation, resulting in an increase of approximately 17.8\% when harmonics up to the $50^{\text{th}}$ order are considered, as highlighted in Figure~\ref{thermal}.
This amplification is not merely a consequence of increased current magnitude, but rather stems from the frequency-dependent growth of the equivalent series resistance (ESR). At elevated harmonic frequencies, skin and proximity effects significantly increase the effective resistance, thereby leading to disproportionate thermal stress and causing localized overheating, accelerated aging, and degradation of power-electronic components, including UPSs and PDUs.

Figure~\ref{IEC60255} illustrates the trip-time characteristics of IEC~60255 IDMT protection relays. The results show that the elevated RMS current induced by GPU power manipulation can readily trigger these protections. For example, under the Standard Inverse curve, a current doubling leads to a trip within approximately two seconds. It manifests as DoS events, abruptly shutting down GPU servers and terminating active workloads. 
In large-scale training or inference, even brief interruptions can cause task failures and cascading restarts, allowing watt-level disturbances to propagate into bit-level service disruptions.

In our exfiltration experiment, binary bits are frequency-shift keying (FSK)-encoded on the GPU load profile using 2~kHz (``1'') and 200~Hz (``0'') modulation frequencies, with a 10~ms bit duration. EMI traces captured via a near-field antenna-coupled USRP B210 reveal that while the absolute spectral power lacks discernible regularity, as shown in the heatmap in Figure~\ref{exfiltration}, the relative energy ratio $M = E_{\text{upper}} / E_{\text{lower}}$ ($E_{\text{lower}}$: 600--800~kHz; $E_{\text{upper}}$: 1080--1180~kHz) remains discriminative. 
By applying a threshold to a 10~ms sliding window, our decoder achieves an accuracy exceeding 99\%. Figure~\ref{exfiltration} demonstrates the successful recovery of a 50-bit test sequence, where every transmitted bit is correctly identified with zero bit errors.
It suggests that the power modulation could potentially serve as a feasible covert channel for clandestine information exfiltration.

\subsection{Countermeasures and Mitigation Strategies}

Mitigating the proposed \texttt{Bit2Watt} attacks requires coordinated defenses across both the cyber and physical layers, as the vulnerability arises from their tight coupling. From the information layer, proactive detection of malicious computational behaviors is essential. This includes identifying abnormal GPU utilization patterns, synchronized workload execution, or anomalous training schedules. Runtime monitoring of power-aware performance counters, combined with anomaly detection, can provide early warning of coordinated power modulation attempts. In addition, integrity verification of training pipelines, model update processes, and job schedulers can reduce the risk of malicious code or adversarial training logic being embedded into large-scale computing workloads.

At the physical layer, enhancing the robustness of data center power infrastructure is equally critical. Local energy buffering mechanisms, such as battery energy storage systems (BESS) and supercapacitors, can effectively absorb fast power fluctuations and high-frequency harmonic energy induced by GPU power modulations \cite{supercapacitor}. Properly sized filtering components and active power conditioning devices can further attenuate harmonic currents and limit their propagation into upstream networks. 

Importantly, these countermeasures should be jointly designed and coordinated across cyber and physical domains. Isolated defenses at either layer are insufficient to address the closed-loop nature of the \texttt{Bit2Watt} risk. A cross-layer security paradigm that integrates computing workload monitoring with power system protection and control is therefore essential to ensure the resilient operation of future data centers and high-renewable power grids, which constitutes future work.

\section{Conclusion}

This paper exposes \texttt{Bit2Watt}, a cyber–physical vulnerability arising from the tight coupling between GPU workloads and renewable-integrated power systems. Coordinated manipulation of legitimate GPU tasks can inject structured power fluctuations, exploiting the negative incremental resistance of constant-power loads and the frequency-shaped impedance of inverter-based DERs. The attack can ultimately degrade power quality, reduce system damping, and potentially trigger cascading failures.
The attack requires no compromise of grid or computing components, operating entirely as a legitimate tenant, and can propagate disturbances back to data centers, causing forced workload interruptions, termed \texttt{Watt2Bit}.
These findings underscore a fundamental shift: as power and computing infrastructures converge, security must be addressed across domains, requiring coordinated defenses that consider workload behavior, power electronics, and grid dynamics.

\vspace{-0.05in}
\section*{Acknowledgments}
\vspace{-0.05in}
The authors are grateful to Runmin Ou and Yiqi Chen for their assistance with the experiments, and to Dr. Chen Yan  of the Ubiquitous System Security Laboratory (USSLab) for his guidance throughout this work. This work was supported by the National Natural Science Foundation of China under Grant 62571474, Grant 62201501 and Grant U25B20188.

\bibliographystyle{alpha}
\bibliography{biblio}

\appendix
\section{Landscape of U.S. Data Centers and Power Grid Stress}

\begin{figure} [h]
	\centering
	\includegraphics[width=0.9\linewidth]{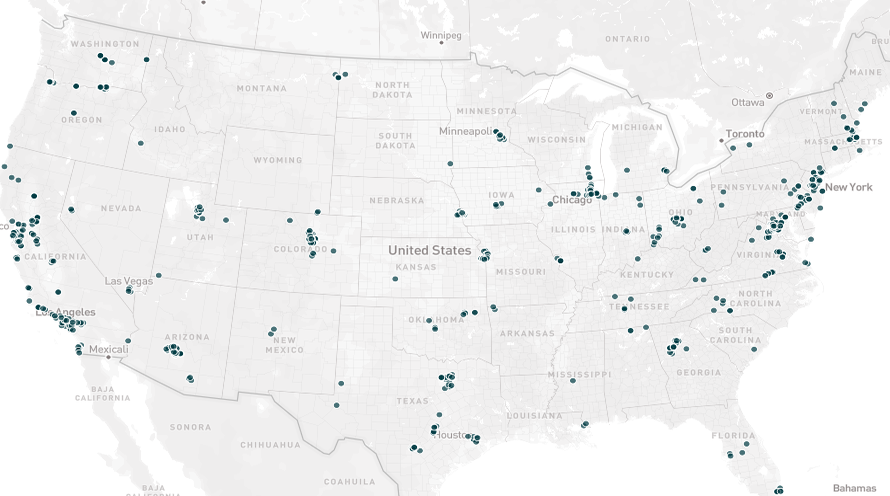} \caption{Spatial distribution of the data centers in the U.S. \cite{USdatacenter}.}
	\label{USdatacenter}
\end{figure}

This appendix provides supplementary evidence for the scale and distribution of data-center loads discussed in Section~\ref{sec:distributionIDC}. 
As shown in Figure~\ref{USdatacenter}, data centers in the U.S. are widely deployed and form dense clusters in several major computing hubs. 
Table~\ref{state_data_centers} further shows that data centers already consume a substantial share of state-level electricity in many regions, reaching 25.6\% in Virginia and exceeding 10\% in several other states. 

These observations indicate that data-center loads are geographically concentrated and large enough to impose significant demand on local power infrastructure. Meanwhile,
GPU clusters exhibit highly dynamic power consumption with abrupt transitions occurring on millisecond timescales. 
When aggregated at scale, these rapid power fluctuations impose concentrated and non-cyclical stress on local grids. Data centers are no longer passive consumers; their reliance on high-precision power delivery renders them highly sensitive to even minor voltage disturbances. As demonstrated by the July 2024 ``Byte Blackout'' event in the PJM Dominion zone \cite{GridStatusByteBlackouts2025}, the spatial concentration of data centers—often referred to as ``Data Center Alleys''—can transform localized transmission faults into regional reliability incidents. Transient events can trigger the near-instantaneous transfer of gigawatt-scale loads to internal UPS systems, producing abrupt load shedding that challenges conventional grid balancing mechanisms.

\begin{table}[t]
\centering
\caption{Share of U.S. State Electricity Used by Data Centers \cite{IDCproportion}}
\label{state_data_centers}
\begin{tabular}{|l|c|l|c|l|c|}
\hline
\multicolumn{2}{|c|}{\textbf{Block 1}} 
& \multicolumn{2}{c|}{\textbf{Block 2}} 
& \multicolumn{2}{c|}{\textbf{Block 3}} \\ \hline

\textbf{State} & \textbf{Share} 
& \textbf{State} & \textbf{Share} 
& \textbf{State} & \textbf{Share} \\ \hline

Virginia & 25.6\% & California & 3.7\% & Connecticut & 1.0\% \\ \hline
Nebraska & 11.7\% & Pennsylvania & 3.2\% & Florida & 0.6\% \\ \hline
Iowa & 11.4\% & New York & 2.8\% & Idaho & 0.6\% \\ \hline
Oregon & 11.4\% & Colorado & 2.7\% & Michigan & 0.5\% \\ \hline
Wyoming & 11.3\% & South Carolina & 2.5\% & South Dakota & 0.5\% \\ \hline
Nevada & 8.7\% & Kentucky & 2.2\% & New Hampshire & 0.2\% \\ \hline
Utah & 7.7\% & Massachusetts & 2.1\% & Maryland & 0.2\% \\ \hline
Arizona & 7.4\% & North Carolina & 1.9\% & Rhode Island & 0.2\% \\ \hline
Washington & 5.7\% & Oklahoma & 1.8\% & Maine & 0.2\% \\ \hline
Illinois & 5.5\% & Alabama & 1.7\% & Wisconsin & 0.2\% \\ \hline
New Jersey & 5.4\% & Ohio & 1.6\% & Indiana & 0.2\% \\ \hline
Texas & 4.6\% & New Mexico & 1.5\% & Louisiana & 0.1\% \\ \hline
North Dakota & 4.4\% & Tennessee & 1.3\% & Hawaii & 0.1\% \\ \hline
Georgia & 4.3\% & Minnesota & 1.2\% & Kansas & 0.0\% \\ \hline
Montana & 3.7\% & Missouri & 1.2\% &  &  \\ \hline

\end{tabular}
\end{table}

\section{Waveform Evidence of Power Adjustment Mechanism of GPU Clusters}

\begin{figure} [h]
	\centering
	\includegraphics[align=t,width=1\linewidth]{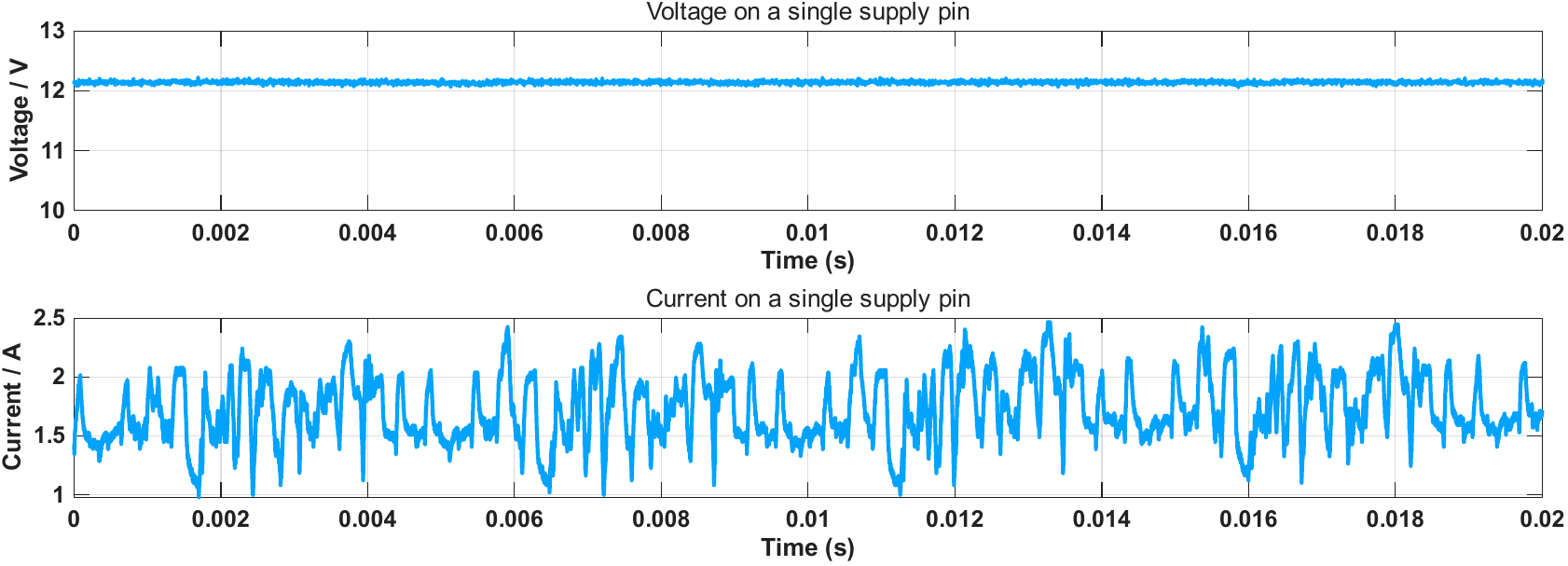}  
	\caption{Board-level voltage and single-line current waveforms of an RTX 3090 GPU during GPT-2 training.}
	\label{GPU_VI}
\end{figure}

\begin{figure} [H]
	\centering
	\includegraphics[align=t,width=1\linewidth]{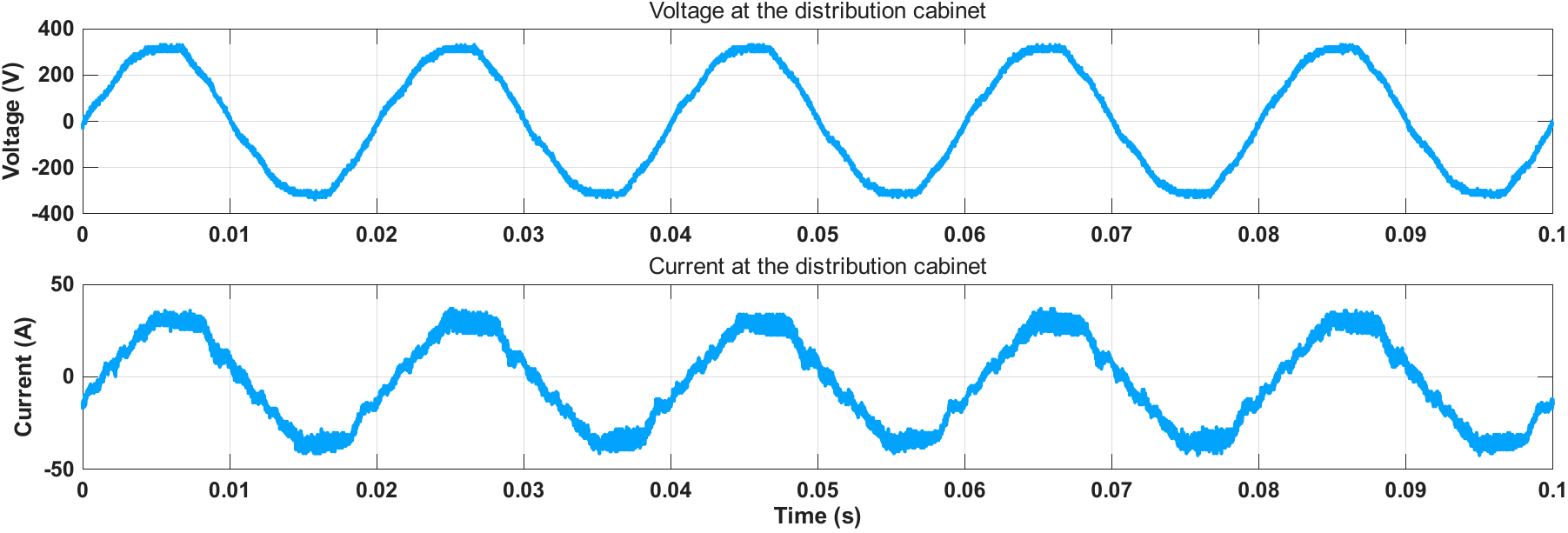}  
	\caption{System-level voltage and current at the distribution cabinet of a GPU cluster.}
	\label{GPU_VI_system}
\end{figure}

This appendix provides supporting evidence for the power adjustment mechanism of GPU clusters discussed in Section~\ref{sec:powermechanism}, using measurements collected under normal operation. 
Figure~\ref{GPU_VI} shows the board-level DC voltage and current measurements of a single GPU, while Figure~\ref{GPU_VI_system} shows the corresponding system-level AC-side waveforms. 
As can be seen, the voltage remains relatively regulated at both levels. However, the board-level DC current shows more pronounced fluctuations and the system-level AC current exhibits pronounced non-sinusoidal behavior, with a heavily distorted waveform and multiple local maxima within each half-cycle.
These measurements qualitatively illustrate that the workload-dependent power variations are primarily realized through changes in the current.

\section{Effect of Timing Jitter on Attack Aggregation}
\label{app:timing_jitter}
\begin{figure} [h]
	\centering
	\includegraphics[align=t,width=0.6\linewidth]{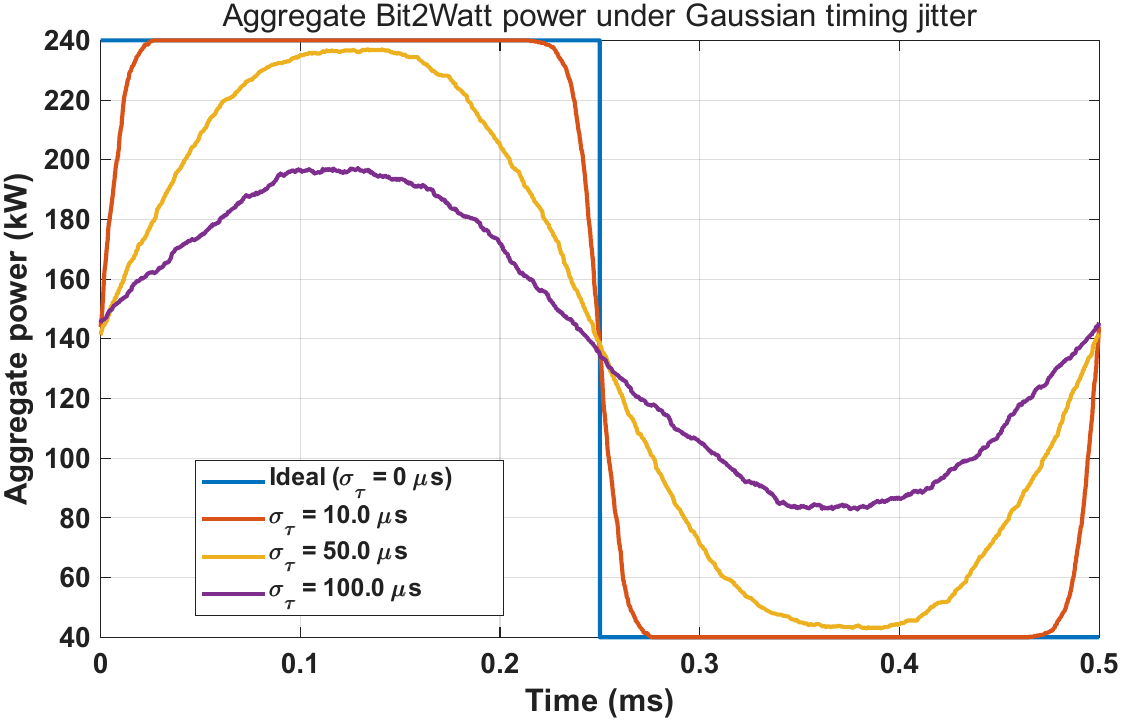}  
	\caption{Aggregated \btw-induced power modulation of 1,000 GPUs under different levels of timing jitter.}
	\label{phaseJitter}
\end{figure}

Figure~\ref{phaseJitter} provides a visual illustration of the timing-jitter effect discussed in Section~\ref{sec:noise}. 
It shows the aggregated \texttt{Bit2Watt} power waveform of 1,000 GPUs under a 2~kHz modulation. 
As $\sigma_\tau$ increases, the imperfect factor progressively smooths the originally square-like waveform and reduces the aggregate modulation amplitude.

\end{document}